\RequirePackage{amsmath}
\documentclass[twocolumn,epjc3]{svjour3}          
\RequirePackage[T1]{fontenc}
\usepackage[absolute]{textpos}
\usepackage{newtxtext}
\usepackage{newtxmath}
\usepackage{graphicx}
\usepackage{flushend}
\usepackage[numbers,sort&compress]{natbib}
\usepackage[colorlinks,citecolor=blue,urlcolor=blue,linkcolor=blue]{hyperref}
\usepackage{breakurl}
\usepackage[caption=false]{subfig}
\usepackage{braket}
\usepackage{amssymb}
\usepackage{enumitem}
\newcommand{\Eqn}[1]{(\ref{#1})}
\newcommand{\MeV}{\,\mbox{MeV}}
\newcommand{\GeV}{\,\mbox{GeV}}
\newcommand{\ppm}{\,\mbox{ppm}}
\newcommand{\MS}{{\overline{\rm MS}}}
\newcommand{\HVP}{\Pi_{\text{had}}}
\newcommand{\M}[2]{\mathcal{M}_{#1}^{(#2)}}
\newcommand{\Amp}[2]{\mathcal{A}_{#1}^{(#2)}}
\newcommand{\A}[3]{\mathcal{A}_{#1,#2}^{(#3)}}

\newcommand{\cA}{{\cal A}}
\renewcommand{\Re}{{\rm Re}}
\newcommand{\had}{\mathrm{h}}
\newcommand{\lep}{\mathrm{lep}}
\journalname{Eur. Phys. J. C}
\begin{document}
\title{Theory for muon-electron scattering @ 10\,ppm\thanksref{t1}}

\subtitle{A report of the MUonE theory initiative}

\author{P.~Banerjee\thanksref{psi}
        \and
        C.~M.~Carloni~Calame\thanksref{pavia} 
        \and
        M.~Chiesa\thanksref{annecy}
        \and
        S.~Di~Vita\thanksref{milano}
        \and
        T.~Engel\thanksref{psi,uzh}
        \and
        M.~Fael\thanksref{kit}
        \and
        S.~Laporta\thanksref{upadua,padua}
        \and
        P.~Mastrolia\thanksref{upadua,padua}
        \and
        G.~Montagna\thanksref{upavia,pavia}
        \and
        O.~Nicrosini\thanksref{pavia}
        \and
        G.~Ossola\thanksref{nyc}
        \and
        M.~Passera\thanksref{padua}
        \and
        F.~Piccinini\thanksref{pavia}
        \and
        A.~Primo\thanksref{uzh}
        \and
        J.~Ronca\thanksref{valencia}
        \and
        A.~Signer\thanksref{e1,psi,uzh}
        \and
        W.~J.~Torres Bobadilla\thanksref{valencia}
        \and
        L.~Trentadue\thanksref{uparma,bicocca}
        \and
        Y.~Ulrich\thanksref{e1,psi,uzh}
        \and
        G.~Venanzoni\thanksref{pisa}
}

\thankstext[$\star$]{t1}{This review is the result of the 2$^\text{nd}$
  WorkStop/ThinkStart that took place 4-7 February 2019 at the
  University of Zurich, as well as the Theory Kickoff Workshop, 4-5
  September 2017, Padova and the MITP Workshop, 19-23 February 2018,
  Mainz.}
\thankstext{e1}{Organisers of the 2$^\text{nd}$
  WorkStop/ThinkStart}

\institute{Paul Scherrer Institut, 5232 Villigen, Switzerland\label{psi}
  \and
  INFN, Sezione di Pavia, 27100 Pavia, Italy\label{pavia}
  \and
  LAPTh, CNRS, 74940 Annecy, France\label{annecy}
  \and
  INFN, Sezione di Milano, 20133 Milano, Italy\label{milano}
  \and
  Physik-Institut, University of Zurich, 8057 Zurich, Switzerland\label{uzh}
  \and
  TTP, Karlsruhe Institute of Technology, 76131 Karlsruhe, Germany\label{kit}
  \and
  Dipartimento di Fisica e Astronomia, University of Padua, 35131
  Padua, Italy\label{upadua}
  \and
  INFN, Sezione di Padova, 35131 Padua, Italy\label{padua}
  \and
  Dipartimento di Fisica, University of Pavia, 27100 Pavia,
  Italy\label{upavia}
  \and
  New York City College of Technology, City University of New York,
  USA\label{nyc}
  \and
  IFIC, University of Valencia -  Consejo Superior de
  Investigaciones Cient\'ificas, 46980 Valencia, Spain\label{valencia}
  \and
  Universit\`{a} di Parma, 43121 Parma, Italy\label{uparma}
  \and
  INFN, Sezione di Milano-Bicocca, 20126 Milano, Italy\label{bicocca}
  \and
  INFN, Sezione di Pisa, 56127 Pisa, Italy\label{pisa}
}

\date{\today}

\maketitle

\begin{abstract}
We review the current status of the theory predictions
for elastic $\mu$-$e$ scattering, describing the recent activities and
future plans of the theory initiative related to the proposed MUonE
experiment.
\end{abstract}
\setlength{\TPHorizModule}{1pt}%
 \setlength{\TPVertModule}{1pt}%
 \begin{textblock}{495}(42,185)%
 \begin{flushright}
    IFIC/20-16\\
    LAPTH-017/20\\
    PSI-PR-20-05\\
    TTP20-018\\
    ZU-TH 11/20
 \end{flushright}
 \end{textblock}%

\section{Introduction} \label{intro}

There is renewed interest in obtaining precise theoretical predictions
for elastic muon-electron scattering. This is to be seen in the
context of MUonE~\cite{Abbiendi:2016xup}, a recent proposal to perform
a very precise measurement of $\mu$-$e$ scattering~\cite{LoI}. A
comparison of experimental data with perturbative calculations can be
used to extract the hadronic vacuum polarisation (HVP) through its
contribution to the running of the QED coupling $\alpha$.  This
follows the original idea of using scattering data to extract the
leading hadronic contribution $a_\mu^\text{HLO}$ to the muon $(g-2)$
from the effective electromagnetic coupling in the space-like
region~\cite{Calame:2015fva}.  The measurement of the running of alpha
in the space-like region from small-angle Bhabha scattering was
proposed in~\cite{Arbuzov:2004wp} and done in~\cite{Abbiendi:2005rx}.

For the planned MUonE experiment, the effect of the HVP changes the
differential cross section of $\mu$-$e$ scattering by up to
$\mathcal{O}(10^{-3})$, depending on the scattering angle of the
outgoing electron. In order to obtain $a_\mu^\text{HLO}$ with a
statistical error similar to current evaluations, the HVP needs to be
extracted from $\mu$-$e$ data with a precision below one percent. Hence,
the accuracy of the total experimental and theoretical error should
not exceed the 10\,ppm level.

The proposal of MUonE is to scatter a $150\GeV$ muon beam on a
Beryllium fixed target. In order to obtain sufficient statistics and
reduce multiple-scattering effects~\cite{Abbiendi:2019qtw}, the target
(about 60\,cm in total) is split into many (about 40) thin layers. The
measurements are done in several stand-alone stations of about 1\,m
length and $10\times 10\,\mbox{cm}^2$ transverse dimension. The
scattering angles of the electron $\theta_e$ and the muon $\theta_\mu$
(in the lab frame) are measured very precisely, but no further
kinematic information is assumed to be available.

From an idealised point of view we thus consider
\begin{align}
  \label{process}
   \mu^\pm(p_1)\, e^-(p_2)\to  \mu^\pm(p_3)\, e^-(p_4) + X
\end{align}
where the initial-state electron is at rest and $X$ stands for any
further radiation. With the energy of the incoming muon set to
$E_1=150\GeV$, the centre-of-mass energy is fixed as $s = m^2+M^2+2 m
E_1 \simeq (400\MeV)^2$, where $m$ and $M$ denote the electron and muon
mass, respectively.  The momentum transfer $t$ ranges from
$t_\mathrm{min} \simeq -(380\MeV)^2$ to zero. Hence, there are two
widely different scales entering the process with $m^2 \ll Q^2$, where
$Q^2$ stands for the large scales $M^2\sim s \sim |t|$. The resulting
large logarithms $\ln(m^2/Q^2)$ will have to be properly accounted for
in the theoretical treatment of the process.

The incoming muon beam consists of either positively or negatively
charged muons and is about 80\% polarised. Since the electrons in the
target are unpolarised and QED is parity conserving, the only effect
of the polarisation is due to the electroweak contributions coming
from the $Z$-boson exchange. At tree level, the latter contributes at
the level of $10\ppm$ and, hence, has to be included.

There are several effects that result in differences from the
idealised process. First, the electrons are bound, and the impact of
bound-state effects should be estimated. Second, there are nuclear
background processes due to $\mu$-$ N$ scattering.
From our point of view, however, the most important aspect of
\Eqn{process} is the selection of elastic scattering. Since photons
are not detected, there is no way of telling how much radiation is
present in $X$. A contribution including $n$ photons results in a
suppression by $\alpha^n$ relative to the leading order (LO), i.e. a
N$^n$LO contribution. Another relevant process is open lepton-pair
production, i.e. $X= e^+\,e^-$ or, albeit with a very small phase
space, $X=\mu^+\,\mu^-$. This amounts to a NNLO QED contribution.
Finally, there is also a background from pion production where the
pion subsequently decays into two photons, i.e. $X = \pi^0 \to
\gamma\,\gamma$. As a last option we mention $X = \pi^+\,\pi^-$, again
with a very small phase space.

In the absence of additional emission $X$ in the final state, i.e.\ for
the elastic $2\to 2$ scattering process, we can derive a simple
functional relation between $\theta_e$ and $\theta_\mu$ that we call
the elasticity curve.  Thus, allowing only events within a small band
around this curve effectively selects nearly elastic events. However,
from a theoretical point of view this is problematic. Making a
stringent cut on the phase space is a further source of large
logarithms, beyond the $\ln(m^2/Q^2)$ mentioned above, that might need
to be resummed.

The precision expected at the MUonE experiment also raises the
question whether possible new physics (NP) could affect its
measurements. This issue was addressed in~\cite{Masiero:2020vxk},
studying possible NP signals in muon-electron collisions at MUonE due
to heavy or light mediators, depending on whether their mass is higher
or lower than ${\cal O} (1\GeV)$. The former were analysed in a
model-independent way via an effective field theory approach, whereas
for the latter the authors discussed scenarios with light spin-0 and
spin-1 bosons. Using existing experimental bounds, they showed that
possible NP effects in muon-electron collisions are expected to lie
below MUonE's sensitivity, therefore concluding that it is very
unlikely that NP contributions will contaminate MUonE's extraction of
the HVP. The authors of~\cite{Dev:2020drf} addressed the sensitivity
of MUonE to new light scalar or vector mediators able to explain the
muon $g-2$ discrepancy. They concluded that the measurement of the HVP
at MUonE is not vulnerable to these NP scenarios.  Therefore, the
analyses of~\cite{Masiero:2020vxk} and~\cite{Dev:2020drf} reach
similar conclusions where they overlap. These results confirm and
reinforce the physics case of the MUonE proposal.

In what follows we will discuss all issues related to obtaining a
theoretical prediction for $\mu$-$e$ scattering at 10\,ppm. We start
in Section~\ref{kinematics} by briefly revisiting the kinematics of
$\mu$-$e$ scattering. Next, we discuss in Section~\ref{fixedorder} the
fixed-order perturbative calculations in QED. This is followed by a
discussion on how to include the HVP in Section~\ref{hvp}. Possible
strategies on how to deal with and estimate the importance of
contributions beyond those included in the fixed-order calculations
are considered in Section~\ref{resummation}. Finally, in
Section~\ref{montecarlo} we give an outlook on how the various pieces
can be combined into a general purpose Monte Carlo code that provides
a sufficiently accurate theoretical prediction, before we present our
summary in Section~\ref{summary}.

\section{Kinematics of \texorpdfstring{$\mu$-$e$}{mu-e} scattering} \label{kinematics}

Let us begin by reviewing for the elastic $\mu$-$e$ scattering process,
\begin{align}
  \mu^\pm(p_1) \, e^-(p_2) \to \mu^\pm(p_3) \, e^-(p_4),
  \label{elprocess}
\end{align}
the basic relations between angles, energies and momenta in the
laboratory frame (LAB) and in the centre-of-mass system (CMS).  In a
fixed-target experiment, where the electron is initially at rest, the
Mandelstam variables $s$ and $t$ are given by
\begin{align}
  &s = M^2 + m^2 +2 m E_1, \notag \\
  &t =  2m^2 -2 m E_4, \notag \\
  &t_\mathrm{min} = - \frac{\lambda(s,M^2,m^2)}{s}  \le t \le 0 .
\end{align}
Here, $E_1$ is the energy of the incident muon, $E_4$ is the electron
recoil energy and
\begin{align}
  \lambda(a,b,c)=a^2+b^2+c^2-2ab -2ac -2bc
\end{align}
is the K\"{a}ll\'{e}n function. The third Mandelstam variable $u$ is
related to $s$ and $t$ in the usual way as $s+t+u=2 M^2 + 2 m^2$.

It is also convenient to define the variable $x$ that is related to
$t$ as
\begin{align}
  x(t)&=\left(1- \sqrt{1-\frac{4 M^2}{t}} \right) \frac{t}{2M^2} &
  \mbox{or} & &
  t(x)&= \frac{x^2\, M^2}{x-1} \, .
\end{align}
With $t_\mathrm{min} \simeq -(380\MeV)^2$ the range of $x$ is $0\leq x
\lesssim 0.933$ and $x=0$ corresponds to $t=0$.

\begin{figure}
  \centering
  \includegraphics[height=5cm]{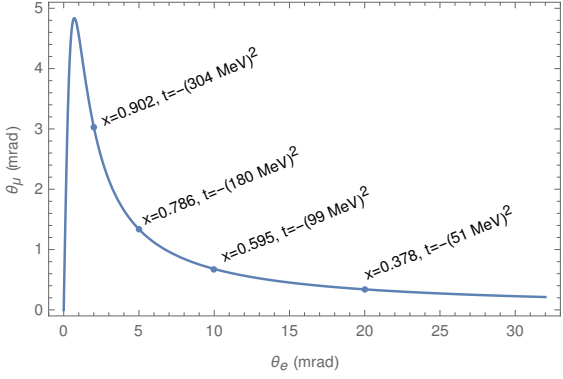}
  \caption{The elasticity curve, i.e. the relation between the muon
  and electron scattering angles for 150~GeV incident muon beam
  momentum.}
  \label{fig:ela}
\end{figure}

The parameters for the Lorentz transformation between the LAB and the
CMS are
\begin{align}
  \gamma &= \frac{E_1+m}{\sqrt{s}} = \frac{s+m^2-M^2}{2m \sqrt{s}},
  \nonumber \\
  \beta &= \frac{|\vec{p}_1|}{E_1+m} = \frac{\lambda^{1/2}(s,M^2,m^2)}{s+m^2-M^2}.
\end{align}
We define the scattering angles $\theta_{e,\mu}$ in the LAB and
$\theta_{e,\mu}^*$ in the CMS as the angles between the direction of
the incident muon and the outgoing electron or muon.  While in the CMS
we trivially have $\theta_e^* = \pi - \theta_\mu^*$, in the LAB frame
the two angles are correlated by the elasticity condition
{\hbadness=10000
\begin{align}
  \tan \theta_\mu =
  \frac{2 \tan \theta_e }{(1+\gamma^2 \tan^2 \theta_e)(1+g_\mu^*)-2},
  \label{eqn:elasticitycondition}
\end{align}
}
where 
\begin{align}
  g_\mu^* = \frac{\beta}{\beta_\mu^*} = \frac{E_1 m+M^2}{E_1 m +m^2}
\end{align}
and $\beta^*_\mu$ is the muon velocity in the CMS.  In the
$\theta_e$-$\theta_\mu$ plane, \Eqn{eqn:elasticitycondition} defines
the elasticity curve depicted in Figure~\ref{fig:ela}. This is the
fundamental constraint for MUonE to discriminate elastic scattering
events from the background of radiative events and inelastic
processes.
Since $g_\mu^* > 1$, the outgoing muon is always emitted in the LAB
forward direction at an angle smaller than $\theta_\mu^\mathrm{max} =
4.8$~mrad (for $E_1=150$~GeV), where
\begin{align}
  \tan \theta_\mu^\mathrm{max} &=
  \frac{1}{\gamma \sqrt{g_\mu^{*2}-1}}, &
  \tan \theta_e \Big|_{\theta_\mu^\mathrm{max}} & =
  \frac{\sqrt{g_\mu^{*2}-1}}{\gamma (g_\mu^{*2}+1)} .
\end{align}
On the contrary, the recoiling electron can be emitted in the whole
LAB forward hemisphere, i.e.\ $0\le \theta_e \le \pi/2$, since $g_e^*
= \beta/\beta_e^*=1$. Therefore, if both scattering angles are below
4.8 mrad there is an ambiguity between muon and electron that must be
resolved by $\mu/e$ discrimination.

The energy and the scattering angle of the electron in the LAB can be
obtained by solving the boost relation $E_4^* = \gamma E_4 - \beta
\gamma \, p_4 \, \cos \theta_e $ for $E_4$. This yields
\begin{align}
  \label{eq:E4}
  \frac{E_4}{m} = 
  \frac{1+\beta^2 \cos^2 \theta_e}{1-\beta^2 \cos^2 \theta_e}.
\end{align}

Going beyond the elastic process \Eqn{elprocess}, by allowing for
additional emission in the final state as described by \Eqn{process},
we have to extend the definitions of the momentum transfer. The
variables
\begin{align}
 t_e &\equiv (p_2-p_4)^2 = 2 m^2 - 2 m E_4 \label{eq:te} \\
 t_\mu &\equiv (p_1-p_3)^2 \label{eq:tm}
\end{align}
now have to be distinguished. Sometimes it is useful to express $t_e$
in terms of the electron scattering angle as
\begin{align}
  t_e = \frac{(2\, m\, \beta \cos\theta_e)^2}{\cos^2\theta_e -1}\,
\end{align}
which follows directly from \eqref{eq:E4} and \eqref{eq:te}.

\begin{figure}
  \centering
  \includegraphics[height=5cm]{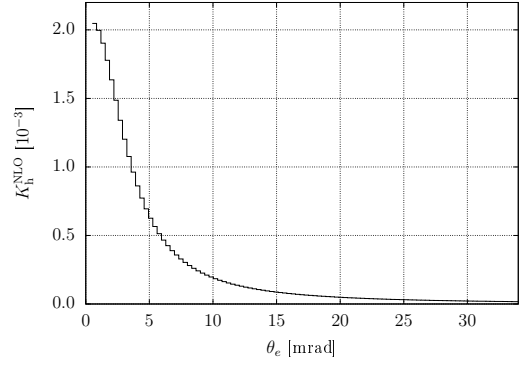}
  \caption{The relative importance of the HVP at NLO in $\mu$-$e$
    scattering as a function of $\theta_e$.}
  \label{fig:hvpnlo}
\end{figure}

Since the contribution of the HVP to $\mu$-$e$ scattering is of central
importance, in Figure~\ref{fig:hvpnlo} we show its leading effect as a
function of the electron scattering angle $\theta_e$. More precisely,
we show the NLO $K$ factor defined as
\begin{align}
	K_\had^{\mathrm{NLO}}(\theta_e) = 
 \frac{d \sigma_\had^{\mathrm{NLO}}}{d\theta_e} / \frac{d \sigma^{(0)}}{d\theta_e},
\label{eqn:RhoHNLO}
\end{align}
where $\sigma^{(0)}$ is the Born cross section and
$\sigma_\had^{\mathrm{NLO}}$ the hadronic contributions at NLO.  As
can be seen from Figure~\ref{fig:hvpnlo} and will be discussed in more
detail in Section~\ref{hvp}, the contribution of the HVP to $\mu$-$e$
scattering is larger for small $\theta_e$, whereas for $\theta_e
\gtrsim 20$~mrad (corresponding to $x\lesssim 0.4$) it is strongly
suppressed.  In Figure~\ref{fig:hvpnlo} the numerical values of the
HVP are from the Fortran library
\texttt{alphaQEDc19}~\cite{Jegerlehner:2001ca, Jegerlehner:2006ju,
Jegerlehner:2011mw, Harlander:2002ur}.

The determination of the HVP will be obtained by a template fit of the
shape of the distribution. For a simplified discussion it is useful to
think in terms of a split into a signal region (small $\theta_e$) and
a normalisation region (large $\theta_e$). In the signal region the
effect is of the order of $10^{-3}$ whereas in the normalisation
region the HVP contribution amounts to $\lesssim 10^{-5}$  and its
error is expected to be below the experimental systematic uncertainty.
A more detailed description of the extraction of the HVP and the
interplay with possible new physics is given at the end of
Section~\ref{hvp:nlo}.

\section{Fixed-order calculations} \label{fixedorder}

In order to achieve our goal of a relative accuracy of 10\ppm, we need
to calculate $\mu$-$e$ scattering at least up to NNLO in the
perturbative expansion in the electromagnetic coupling $\alpha \sim
1/137$. In addition, we need a flexible setup that allows for the
computation of arbitrary infrared safe observables, i.e. a parton
level Monte Carlo (MC). The latter aspects will be discussed in
Section~\ref{montecarlo}. In this section we discuss the main features
of the analytic fixed-order computations. We stress that by LO, NLO,
and NNLO we imply a strict fixed-order expansion in the on-shell
coupling $\alpha$, without any resummation whatsoever. Issues related
to resummation will be discussed in Section~\ref{resummation}.

\subsection{Leading order}

\begin{figure}
\centering
\subfloat[LO QED diagram\label{fig:lo:qed}]{
    \includegraphics[width=0.4\columnwidth]{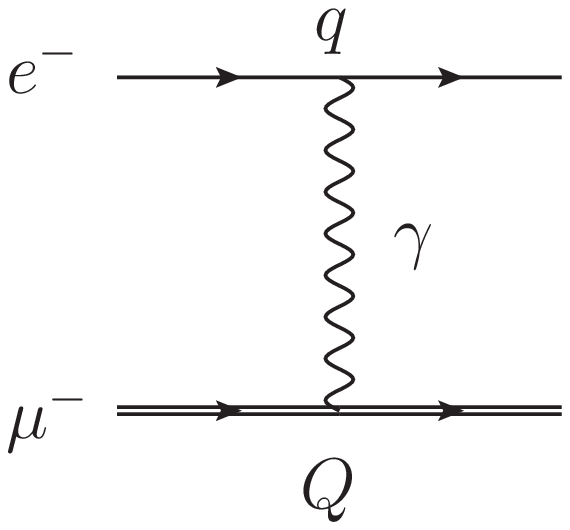}}
\\
\subfloat[HVP contribution at NLO\label{fig:lo:hvp}]{
    \includegraphics[width=0.35\columnwidth]{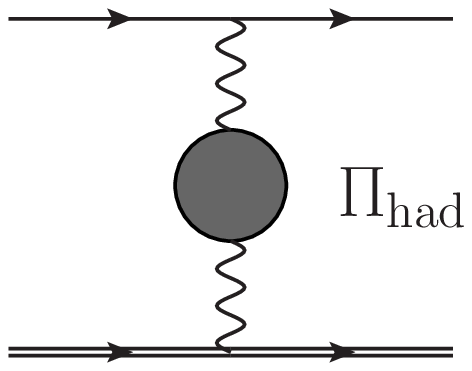}}
\qquad
\subfloat[LO $Z$-boson exchange\label{fig:lo:z}]{
    \includegraphics[width=0.35\columnwidth]{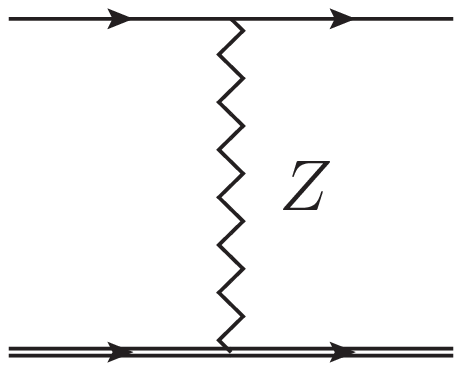}}
\caption{LO contributions from QED, HVP and the $Z$-boson
  exchange}\label{fig:lo}
\end{figure}

Starting at LO in QED there is a single diagram with a $t$-channel
exchange of a photon. It is precisely this feature that makes this
process ideal to extract the HVP. The dominant contribution of the HVP
is simply given by the insertion of the hadronic bubble $\HVP$ into
the photon propagator, as shown in Figure~\ref{fig:lo:hvp}. It is
precisely the effect of this contribution that is shown in
Figure~\ref{fig:hvpnlo}.

As indicated in Figure~\ref{fig:lo} we often (formally) distinguish
the charges of the electron and muon. Denoting them by $q$ and $Q$
respectively, the LO amplitude can be written as
\begin{align}
  \label{AMPlo}
  \cA^{(0)}(\mu e\to\mu e) \equiv \Amp{n}{0}  = q Q\, \A110\, ,
\end{align}
where the superscript indicates the number of loops. The two integer
subscripts of the last expression indicate the power of $q$ and $Q$.
The two-particle final state is indicated by the subscript $n$ of the
first expression, where $n=2$ is implicitly understood. To obtain a
(differential) LO cross section $d\sigma^{(0)}$ we simply integrate
the (squared) matrix element $\M{n}{0}$ over the two-particle phase
space
\begin{align}
  \label{Mlo}
   d\sigma^{(0)} &= \int d\Phi_n \M{n}{0} 
    = \int d\Phi_n \big| \Amp{n}{0} \big|^2\, ,
\end{align}
where cuts applied by the experiment and the definition of the
observable are understood. The leading-order differential cross
section is given by
\begin{align}
  \frac{d\sigma^{(0)}}{d t} = 4 \pi \alpha^2
  \frac{(M^2+m^2)^2 - s u + t^2/2}{t^2\, \lambda(s,M^2,m^2)}\, .
\end{align}
Because $\M n0\sim 1/t^2$ and, hence, $d\sigma^{(0)}/(d t)\sim 1/t^2$
the total cross section is not well-defined. Therefore, we always have
to apply cuts to the integration to avoid the region $t\sim 0$.

At LO (and NLO), effects due to the electron mass $m$ are suppressed by
$z^2$ where
\begin{align}
  \label{def:z}
  z \equiv \frac{m}{M}\, .
\end{align}
Hence, they have to be taken into account at -- and even beyond -- LO
to achieve a 10\ppm\ prediction.

The contributions due to the exchange of a $Z$ boson are strongly
suppressed because of its large mass $M_Z$. However, the interference
between the $Z$-boson and photon-exchange diagrams is suppressed with
respect to the LO QED contribution only by $Q^2/M_Z^2 \simeq
10^{-5}$. Hence, this effect is relevant and needs to be taken into
account in the calculation.

\subsection{Next-to-leading order}

Going to NLO, the separately divergent real and virtual contributions
have to be combined to obtain a physical result. Following earlier
efforts~\cite{Bardin:1997nc, Kaiser:2010zz}, recently a fully
differential NLO code~\cite{Alacevich:2018vez} has been used to
perform a detailed phenomenological study, taking into account the
full $m$ dependence. The NLO contributions can be split into gauge
invariant parts by separating the contributions into powers of $q$ and
$Q$. Thus we decompose the NLO amplitude $\Amp{n}{1} \equiv
\cA^{(1)}(\mu e\to \mu e)$ as
\begin{align}
  \label{AMPnlo}
  \Amp{n}{1} &= q^3 Q\, \A311 + q Q^3\, \A131 + q^2 Q^2\,\A221\, .
\end{align}
The leptonic vacuum polarisation contributions are part of the full
NLO calculation. However, we sometimes treat them separately as they
are somewhat closer connected to the signal extraction.

\begin{figure} \begin{center}
\subfloat[Example for a contribution to the squared matrix element
  $\propto q^4Q^2$]{
  \includegraphics[width=0.3\textwidth]{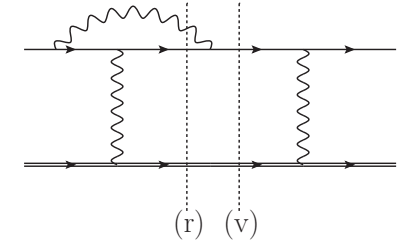}} \qquad
\subfloat[Example for a contribution to the squared matrix element
  $\propto q^3Q^3$]{
  \includegraphics[width=0.3\textwidth]{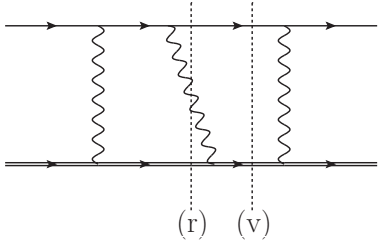}} \qquad
\end{center}
\caption{Examples of NLO QED contributions to $\M{n+1}{0}$ (r), and
$\M{n}{1}$ (v). Analogous muonic contributions proportional to $q^2
Q^4$ are implied. \label{fig:nlo}}
\end{figure}

The virtual corrections are then obtained by integrating over the
$n=2$ parton phase space the renormalised squared matrix element
\begin{align}
  \label{AMPv}
  \M{n}{1} &= 2\,\Re\big[ \Amp{n}{1}\times (\Amp{n}{0})^* \big]\, .
\end{align}
Typically, the masses and wave functions are renormalised in the
on-shell scheme, whereas for the coupling, either the on-shell scheme
or the $\MS$-scheme can be used.  Similarly, the real corrections are
obtained by integrating over the $n+1=3$ parton phase space the
squared matrix element
{\hbadness=10000
\begin{align}
  \label{AMPr}
  \M{n+1}{0} &=  \big| \Amp{n+1}{0} \big|^2\,
\end{align}}
where the amplitude $\Amp{n+1}{0} = \cA^{(0)}(\mu e\to \mu e \gamma)$
is also decomposed  according to
\begin{align}
  \label{AMPg}
  \Amp{n+1}{0} &= q^2 Q \A210 + q Q^2 \A120\, .
\end{align}
The cross section is obtained as the sum 
\begin{align}
  \label{Snlo}
   d\sigma^{(1)} &= d\sigma^{(\mathrm{v})} + d\sigma^{(\mathrm{r})} \\
   &=  \int d\Phi_n \M{n}{1} + \int d\Phi_{n+1} \M{n+1}{0}\, .
   \nonumber
\end{align}
As illustrated in Figure~\ref{fig:nlo}, the terms $\sim q^4 Q^2$
($\sim q^2 Q^4$) in $\M{n}{1}$ and $\M{n+1}{0}$ correspond to the
corrections due to photon emission from the electron (muon) and are
hence called electronic (muonic) contribution. There are also mixed
terms $\sim q^3 Q^3$. The latter flip the sign if $\mu^-$ is changed
to $\mu^+$. As will be discussed in Section~\ref{montecarlo},
electronic effects are actually dominant~\cite{Alacevich:2018vez}.
Regarding the virtual corrections, the electronic contribution only
requires $\A311$ which is simpler to compute than $\A221$. Such
considerations become more important when discussing NNLO
contributions.

Keeping a finite $m$ complicates the computation of the virtual
corrections. On the other hand, it serves as a regulator for collinear
singularities which are replaced by $\log(m^2/Q^2)$ and only soft
singularities are left.  In~\cite{Alacevich:2018vez} the latter are
regularised using a photon mass. There are two additional independent
parton level Monte Carlo codes~\cite{NLOmfmp, NLOus} using dimensional
regularisation for IR singularities. These codes have been compared
to~\cite{Alacevich:2018vez} and full agreement has been found.

Electroweak (EW) NLO corrections are not expected to be required at
the 10\ppm\ level. This was explicitly verified
in~\cite{Alacevich:2018vez}.

\subsection{Next-to-next-to-leading order}

A complete result for NNLO QED corrections to $\mu$-$e$ scattering is
not yet available. However, there are already several partial results
and a large theoretical effort is under way to complete the full NNLO
calculation.

\begin{figure}
\begin{center}
\subfloat[Example for a contribution to the squared matrix element
  $\propto q^6Q^2$\label{fig:nnlo:el}]{
  \includegraphics[width=0.32\textwidth]{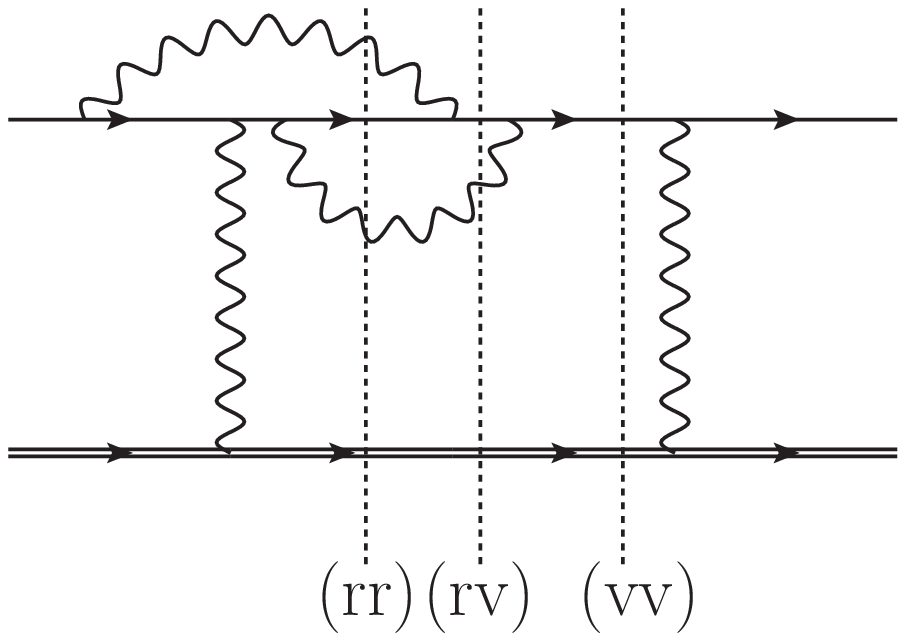}} \qquad
\subfloat[Example for a contribution to the squared matrix element
  $\propto q^5Q^3$\label{fig:nnlo:mix}]{
  \includegraphics[width=0.32\textwidth]{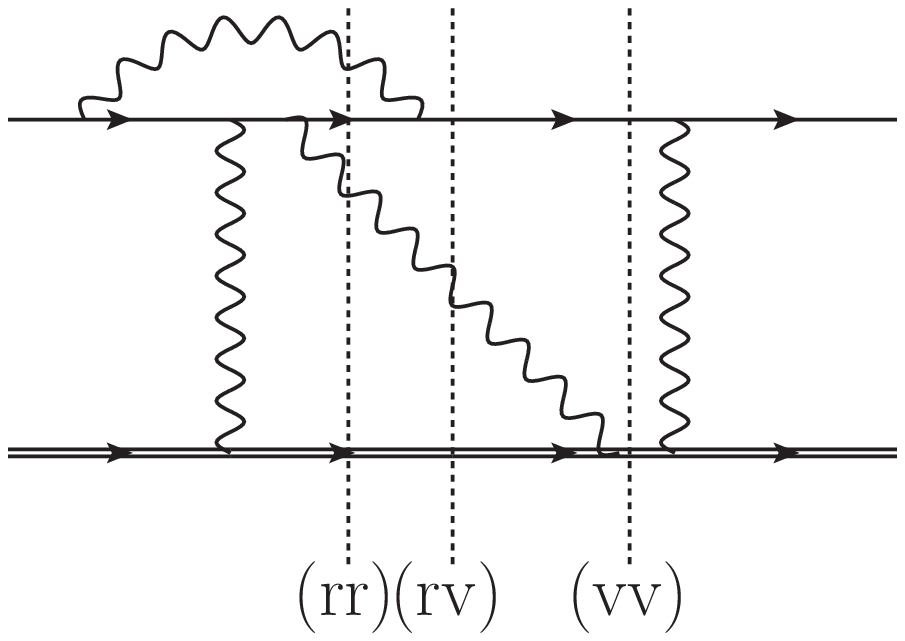}} \qquad
\end{center}
\caption{Examples of NNLO QED contributions to $\M{n+2}{0}$ (rr),
  $\M{n+1}{1}$ (rv), and $\M{n}{2}$ (vv). Analogous contributions with
  $q^l Q^{8-l}$ with $l\in\{4,3,2\}$ are understood. \label{fig:nnlo}}
\vspace{-2mm}
\end{figure}

Following the notation of \Eqn{AMPlo}, \Eqn{AMPnlo} and
\Eqn{AMPg} and using $\cA_{n+2} = \cA(\mu e\to \mu e \gamma\gamma)$,
the required amplitudes for the NNLO corrections are
\begin{align}
\label{AMPnnlo}
\Amp{n}{2} &= q^5 Q\, \A512 + q^4 Q^2\, \A422
  + q^3 Q^3\,  \A332 \\
  &+ q^2 Q^4\, \A242 + q Q^5\, \A152 \nonumber\\[5pt] 
\label{AMPgnlo}
\Amp{n+1}{1} &= q^4 Q \A411 + q^3 Q^2 \A321 \\
 &+ q^2 Q^3 \A231 + q Q^4 \A141
 \nonumber \\[5pt]  
\label{AMPgg}
  \Amp{n+2}{0} &=  q^3 Q \A310 +  q^2 Q^2 \A220 +  q Q^2 \A130
\end{align}
Similarly, for  the matrix elements we need
\begin{align}
\label{AMPvv}
  \M{n}{2} &= 2\,\Re\Big[ \Amp{n}{2} \times (\Amp{n}{0})^* \big]
           +  \big| \Amp{n}{1} \big|^2
\\[3pt]
\label{AMPvr}
  \M{n+1}{1} &=  2\,\Re\big[ \Amp{n+1}{1}\times (\Amp{n+1}{0})^* \big]
\\[3pt]
\label{AMPrr}
  \M{n+2}{0} &= \big| \Amp{n+2}{0} \big|^2
\end{align}
for the double-virtual (vv), real-virtual (rv) and double-real (rr)
corrections. They have to be integrated over the $n=2$, $n+1=3$ and
$n+2=4$ parton phase space, respectively,
\begin{align}
  \label{Snnlo}
  d\sigma^{(2)} &= d\sigma^{(\mathrm{vv})} + d\sigma^{(\mathrm{rv})}
   + d\sigma^{(\mathrm{rr})}  \\
    &=  \int d\Phi_n \M{n}{2} + \int d\Phi_{n+1} \M{n+1}{1}
     + \int d\Phi_{n+2} \M{n+2}{0}\, .
  \nonumber
\end{align}
The interplay between these three parts is illustrated in
Figure~\ref{fig:nnlo} where different cuts to the same diagram squared
represent contributions to $d\sigma^{(\mathrm{rr})}$,
$d\sigma^{(\mathrm{rv})}$ and $d\sigma^{(\mathrm{vv})}$, respectively.
From a theory point of view, there is a choice whether to include the
sub-process $\mu e\to\mu e+ee$ in $\M{n+2}0$. Assuming $m>0$, this is
a separate IR finite contribution.

The main bottleneck for a NNLO calculation keeping the full $m$
dependence is the evaluation of the two-loop amplitude
$\Amp{n}{2}(m)$. It is not clear if a complete NNLO calculation with
full $m$ dependence is feasible in the next years. Fortunately, this
is also not really required. The electronic contributions can be
computed with full $m$ dependence. For the remaining contributions, an
approximate treatment for the NNLO corrections, i.e. an expansion in
$z$ is expected to be sufficient to obtain 10\ppm\ precision in the
theoretical prediction.

Usually we refrain from listing the dependencies of the amplitudes on
the momenta and the masses, $m$ and $M$. However, sometimes we will
have to indicate how we treat the dependence on the electron
mass. Either we keep it completely as in $\Amp{n}{2}(m)$, or we set it
to zero as in $\Amp{n}{2}(0)$. A third option is to consider an
expansion in $m$, using $m^2\ll \{M^2,Q^2\}$. We will indicate this in
the notation by writing $\Amp{n}{2}(z)$ having in mind $m=z M$ with
$z\ll 1$.

Whether or not we keep the electron mass will alter the form of how
the IR singularities of $\M{n}{2}$ manifest themselves. Using
dimensional regularisation with $d=4-2\epsilon$, the highest pole of
$\M{n}{2}(0)$ is $1/\epsilon^4$ which corresponds to
double-soft-collinear poles. On the other hand, $\M{n}{2}(m)$ and
$\M{n}{2}(z)$ will only have $1/\epsilon^2$ poles. The
double-soft-collinear poles are now replaced by $1/\epsilon^2\,
\log^2(z)$.

A similar change happens in the real-virtual and double-real
contribution. Introducing an electron mass regularises the collinear
singularities in the phase-space integration, again transforming the
corresponding $1/\epsilon$ poles to $\log(z)$ terms. Of course, for a
physical cross section, all final-state collinear (and soft)
singularities cancel. Thus, as for all regularisation procedures, a
cross section is independent of which regularisation is chosen for the
collinear singularities. The difference between $d\sigma^{(2)}(m)$ and
$d\sigma^{(2)}(0)$ or $d\sigma^{(2)}(z)$ is in terms of the form $z^p
\log^l(z)$ that are finite (and actually vanish) for $m\to 0$.  An
advantage of the regularisation with a finite electron mass is that
the initial-state collinear logarithms are manifest in the fixed-order
contributions.

In what follows we will now consider how to obtain a sufficiently
precise approximation to a complete NNLO calculation, comparing
different approaches on how to treat the electron mass.

\subsubsection*{Massive electron}
We start by noting that also at NNLO the electronic emission is the
dominant contribution. This corresponds to the terms $\sim q^6 Q^2$ in
$d\sigma^{(2)}$, with an example shown in Figure~\ref{fig:nnlo:el}.
This part can actually be computed with full $m$ dependence. It is a
problem with only one active mass scale, $m$, in the loops and the
two-loop virtual corrections $2\,\Re[\A512\times(\Amp{n}{0})^*]$ can
be obtained from the heavy-quark (actually lepton) two-loop form
factor~\cite{Bonciani:2003ai, Bernreuther:2004ih}. For the moment we
do not include the HVP insertion in the two-loop diagrams. This will
be dealt with in Section~\ref{hvp}.

In order to combine this with the double-real and real-virtual
contributions using dimensional regularisation, a suitable NNLO
subtraction scheme has to be implemented. One example of such a scheme
is the FKS$^2$~\cite{Engel:2019nfw} which extends the NLO FKS
subtraction scheme~\cite{Frixione:1995ms,Frederix:2009yq} to NNLO in
the case of massive QED where only soft singularities are present.
FKS$^2$ was successfully tested for the muon decay with full electron
mass.  Preliminary results for the $q^6Q^2$ terms have recently been
presented in~\cite{Engel:2019,Ulrich:2019}.  Of course, it is trivial
to adapt these computations for the purely muonic emission, i.e. the
terms $\sim q^2 Q^6$. But they are expected to be numerically much
less important. A similar calculation has been done in the context of
lepton-proton scattering in~\cite{Bucoveanu:2018soy}, where the
electronic terms for $e\,p\to e\,p$ scattering have been computed
using a phase-space slicing method.

Contributions where both emission from the electron and muon line are
involved are technically much more challenging. An example is shown on
Figure~\ref{fig:nnlo:mix}. In fact, an analytic computation of the
two-loop amplitude, keeping the full $m$ dependence, is probably not
feasible in the near future. However, the reduction to master
integrals of $\M{n}{2}(m)$ is currently under investigation. This
could be combined with a numerical evaluation of the master integrals.
Another approach would be a completely numerical evaluation of the
amplitude, even avoiding a reduction to master integrals. Even if a
complete result for $\M{n}{2}(m)$, suitable for a Monte Carlo code, is
not expected to be available within the next few years, these efforts
are extremely useful as cross checks for other approaches (see below).

\begin{figure}[t]
\centering
\includegraphics[width=0.45\textwidth]{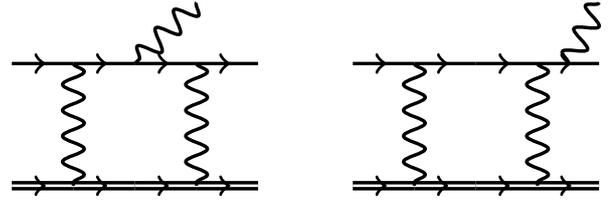}
\caption{Examples of Feynman diagrams $\sim q^3 Q^2$ contributing to
  the real-virtual corrections to $\mu$-$e$ scattering at NNLO in
  QED. Related diagrams $\sim q^2 Q^3$, where the real photon is
  radiated from a muonic line, can be obtained from these by means of
  symmetries.}
\label{fig:diag5pt1L}
\end{figure}

Apart from the two-loop amplitude also the full real-virtual
corrections $\M{n+1}{1}(m)$ need to be computed, i.e. the interference
between one-loop and the Born amplitude of $e\mu\to e\mu\gamma$. Even
though their integration over the $n+1=3$ phase space entails an IR
(soft) singularity, the order $\epsilon$ terms of $\M{n+1}{1}(m)$ are
not really required, if a suitable subtraction scheme such as FKS$^2$
is used.
{\hbadness=10000

}
The calculation of these interference terms was considered
in~\cite{Dondi:2019th}, where both cases with massless and massive
electrons were studied.  The real-virtual contributions are
encompassed in the $\Amp{n+1}{1}$ term, and comprise 44 Feynman
diagrams which are generated with the \textsc{Mathematica} packages
\textsc{FeynArts} and \textsc{FeynCalc}~\cite{Hahn:2000kx,
Shtabovenko:2016sxi}.  Four of these diagrams do not actually
contribute since they automatically cancel out at integrand level
because of Furry's theorem, while the remaining ones can be split into
two sets: \textit{i)} where the real photon emission occurs from a
muonic internal or external line and \textit{ii)} where the emission
is from an electron line.  When both leptons are massive the
contributions from the two sets can be related via symmetries, namely
exchanging the electron and muon masses and charges as well as the
respective external momenta. This fact was exploited to halve the
number of diagrams to be evaluated to just 20. Representative diagrams
are depicted in Figure~\ref{fig:diag5pt1L}.

These unrenormalised amplitudes are then inserted in the real-virtual
interference term~\eqref{AMPrr}, which is computed with massive
electrons to assess the impact of the massless electron approximation
enforced for the calculation of other contributions. The first steps
towards a fully-analytic evaluation of this contribution were
undertaken, using the same automatic framework employed for the
calculation of $\mathcal{M}_{n}^{\left(2\right)}$. The integrands
depend on the two mass parameters plus five kinematic variables, which
were parametrised using the \textit{Momentum Twistor
  parametrisation}~\cite{Hodges:2009hk, Badger:2016uuq,
  Driencourt-Mangin:2019yhu} in preparation for the adaptive integrand
decomposition in \textsc{Aida}. Subsequently the amplitudes were
simplified via integration-by-parts identities~\cite{Tkachov:1981wb,
  Chetyrkin:1981qh, Laporta:2001dd} generated with the
package \textsc{Kira}~\cite{Maierhoefer:2017hyi}, identifying 45
master integrals.

The interferences were then matched with counterterm amplitudes
generated in \textsc{FeynCalc}, employing the on-shell renormalisation
scheme. The cancellation of the leading ultraviolet (UV) poles in
dimensional regularisation in the re\-nor\-ma\-lised amplitudes was verified
numerically.

\subsubsection*{Massless electron}
Neglecting the electron mass reduces the difficulty of the problem
from extreme to very high.  Fortunately, there has been an impressive
effort devoted to this computation, such that the evaluation of the
two-loop amplitude for massless electrons, $\Amp{n}{2}(0)$, is close
to completion.

\begin{figure}[t]
\centering
\begin{center}
\vspace{0.7cm}
\vspace{4.8mm}
\includegraphics[width=0.45\textwidth]{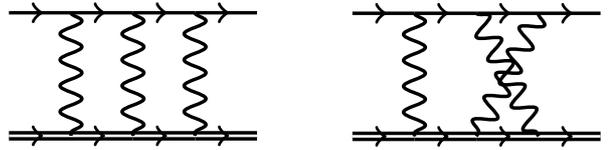}
\vspace{4.6mm}
\caption{Representative examples of two-loop diagrams contributing to
  $\mu$-$e$ scattering at NNLO in QED.}\label{fig:NNLOdiags}
\end{center}

\end{figure}

The amplitude $\Amp{n}{2}(0)$ receives contributions from 69 Feynman
diagrams, which are generated with the help of the packages {\sc
  FeynArts/FeynCalc}~\cite{Hahn:2000kx,Shtabovenko:2016sxi} and its
evaluation requires the calculation of ${\cal O}( 10^4)$
integrals. Owing to the use of adaptive integrand
decomposition~\cite{Mastrolia:2016dhn, Mastrolia:2016czu}, implemented
in the in-house package {\sc Aida}~\cite{AIDA}, and
integration-by-parts identities~\cite{Tkachov:1981wb,
  Chetyrkin:1981qh, Laporta:2001dd}, implemented in the public
routines {\sc Reduze}~\cite{vonManteuffel:2012np,
  vonManteuffel:2014qoa} and {\sc Kira}~\cite{Maierhoefer:2017hyi},
the amplitude -- to be precise, the interference term $2\,\Re\big[
  \Amp{n}{2} \times (\Amp{n}{0})^* \big]$ -- has been simplified, and
written as a linear combination of an integral basis formed by about
120 elements~\cite{Mastrolia:2017pfy, DiVita:2018nnh}, dubbed master
integrals.  The latter have been successfully evaluated by means of
the differential equation technique~\cite{Barucchi:1973zm,
  Kotikov:1990kg, Remiddi:1997ny, Gehrmann:1999as, Henn:2013pwa} in
combination with the Magnus exponential method~\cite{Argeri:2014qva,
  DiVita:2014pza}. Originally evaluated in a non-physical region,
where the mathematical complexity was found to be more limited, the
analytic evaluation of the master integrals to the physical scattering
region was recently obtained~\cite{DiVita:2019lpl}.  The analytic
expressions of the master integrals, numerically evaluated with the
help of {\sc GiNaC}~\cite{Bauer:2000cp}, were successfully validated
against the numerical values provided either by {\sc
  SecDec}~\cite{Borowka:2015mxa} or, for the most complicated
integrals, coming from the 7-propagator graphs, by an in-house
algorithm. Representative diagrams are depicted in
Figure~\ref{fig:NNLOdiags}.

\begin{figure}
\centering 
\includegraphics[width=0.9\columnwidth]{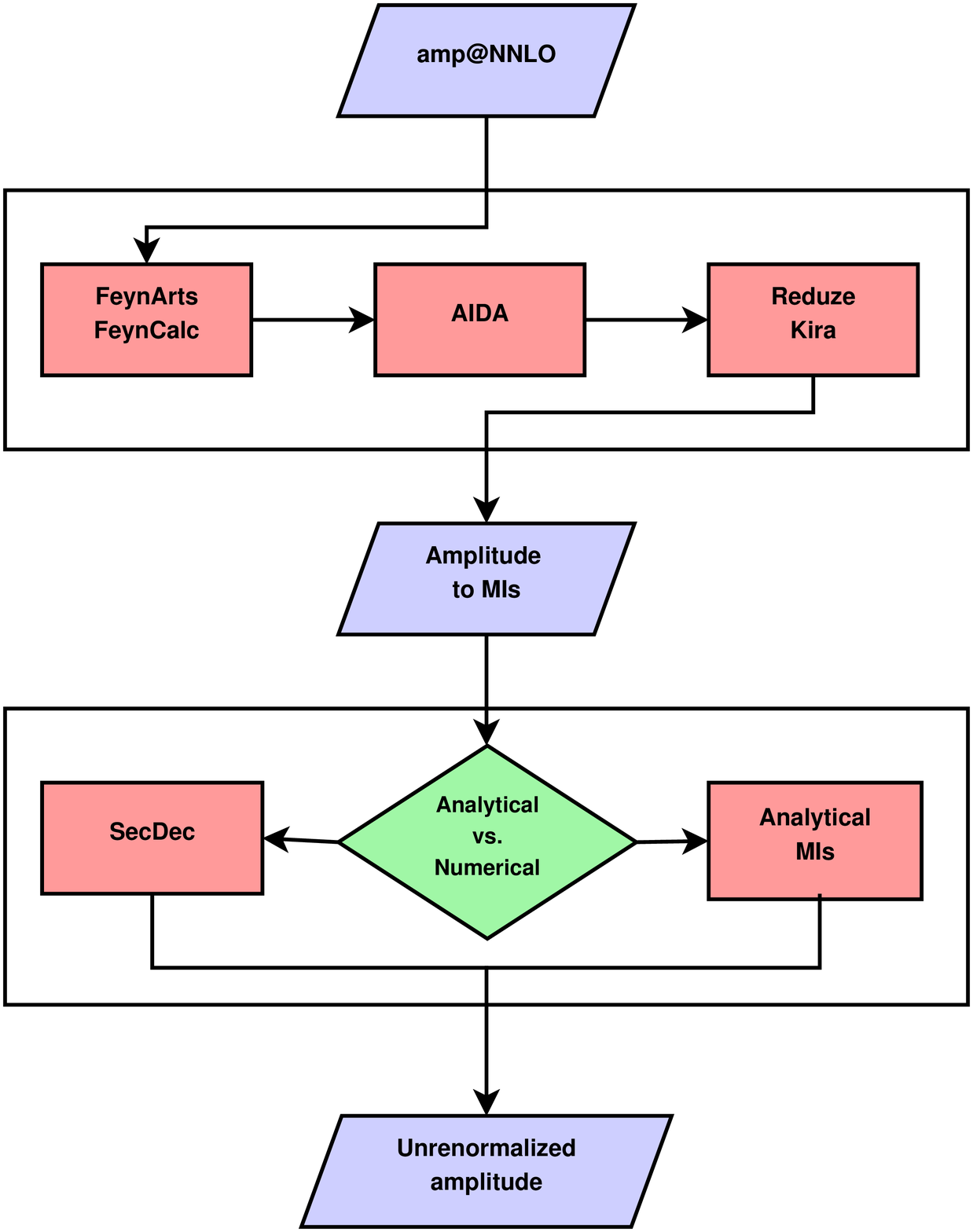}
\caption{Flowchart of the algorithm to evaluate the two-loop
  amplitude. }\label{fig:NNLOflowchart}
\end{figure}

The non-trivial evaluation of the (unrenormalised) two-loop amplitude
for the $\mu$-$e$ scattering required the development of a high-level
automated tool, exploiting the synergy of different packages embedded
in a {\sc Mathematica} framework, whose flowchart is depicted in
Figure~\ref{fig:NNLOflowchart}.  UV divergences arising from
divergent loop integrals are regularised within the dimensional
regularisation scheme, to be later removed by means of a diagrammatic
approach to the renormalisation. In particular, the counterterm
Lagrangian provides additional Feynman rules, which can be adopted in
our automatic framework for generating the additional diagrams,
yielding a UV finite amplitude. Given the masslessness of the
electron, we adopted an {\it hybrid} renormalisation scheme choice:
\begin{itemize}
\item
    $\MS$ scheme for the \emph{coupling};
\item
    on-shell scheme for the \emph{muon} mass.
\end{itemize} 
For a recent review on the state of the calculation, see
also~\cite{Ronca:2019kcw}.

In principle, the full computation can be done with massless
electrons. In the phase-space integration, this results in collinear
singularities. Hence the subtraction (or any other) scheme used will
need to be adapted to this case. Unfortunately this will destroy the
simple divergence structure of massive QED that was exploited in
FKS$^2$.

There is one further subtlety when performing the calculation with
massless electrons everywhere. While final-state singularities will
cancel for $m=0$ in any sufficiently inclusive
observable~\cite{Kinoshita:KLN}, the same cannot be said about
initial-state collinear singularities\footnote{Such poles exist even
  though the initial-state electron is at rest as the total cross
  section is Lorentz invariant.}. The corresponding $\epsilon$ poles
will remain unless properly treated. There are multiple somewhat
related ways to make these expressions well defined such as the
Weizsacker-Williams approach, the structure function
approach~\cite{Kuraev:1985hb,Ellis:166310} or the QED parton
distribution function approach~\cite{Frixione:2019lga}. These
techniques where honed in the LEP era and will work at the required
accuracy.

A final problem with a purely massless calculation is the
restrictions imposed by the phrase `sufficiently inclusive observable'
of the KLN theorem~\cite{Kinoshita:KLN}. This will make a quantity
such as $\theta_e$ inaccessible without breaking IR safety or defining
a jet-like observable.

\subsubsection*{Massified electron}
Given that the electron mass is a natural cutoff for collinear
emission, it seems to be natural to use $m$ as a collinear
regulator. Apart from reducing the complexity of the IR subtraction
for the real integration, this will also facilitate the combination of
a fixed-order result with parton-shower Monte Carlo codes and
automatically produce the $\log(m)$ terms that are present in
distributions. 

In order to do this, we will have to massify $\M{n}{2}(0)$,
i.e. transform it into $\M{n}{2}(z)$ that contains the leading
logarithmic terms $\log(m)$. Initially, this problem has been
addressed for Bhabha scattering~\cite{Penin:2005kf, Penin:2005eh} and
then been generalised~\cite{Mitov:2006xs, Becher:2007cu} using a
factorisation approach. A further generalisation is needed if two
different non-vanishing masses exist, as in our case $M$ and $m$. This
has been studied in the context of the muon decay~\cite{Engel:2018fsb}
and will allow to obtain $\M{n}{2}(z)$ from $\M{n}{2}(0)$.

To achieve this, an approach based on soft collinear effective theory
(for an introduction to SCET see~\cite{Becher:2014oda}) and the method
of regions~\cite{Beneke:1997zp} is used. Loop integrals contributing
to $\M{n}{2}(m)$ are expanded in $z$ by taking into account all
relevant scalings of the loop momenta $k_i$ (regions) and expand the
integrand in all these regions. After expansion of the integrand, the
integrations are simplified and adding up all contributions reproduce
the expansion of the full integral.  In our case, the relevant regions
are hard $k_i\sim(1,1,1)$, soft $k_i\sim(z,z,z)$ and, ultrasoft
$k_i\sim (z^2,z^2,z^2)$. Further, we need collinear $k_i\sim(z^2,1,z)$
for the in-coming and anti-collinear scaling $k_i\sim(1,z^2,z)$ for
the out-going electron. Here we have used light-cone coordinates
$k_i=(k_+,k_-,k_\perp)$.  Each external electron defines a collinear
direction (either the one of $k_-$ or of $k_+$) that has to be taken
into account.

In principle, this can be done to any power in $z$. However,
restricting ourselves to the leading power, the matrix element
factorises as
\begin{align}
  \label{factorizemM}
  \M{n}{2}(z) =\prod_{i=1,2} \sqrt{Z_i^{\phantom{\prime}}(m)}
        \times{\cal S}\times \M{n}{2}(0)\, .
\end{align}
The hard contributions correspond to the massless matrix element
$\M{n}{2}(0)$ the computation of which was discussed above. The soft
part, ${\cal S}$, is also process dependent and will have to be
computed for $\mu$-$e$ scattering. However, the computation of the soft
part is much simpler than the full amplitude. It obtains contributions
from fermion-loop diagrams and can be tested against a fully massive
computation of the fermion loops~\cite{Fael:2019nsf}.  The collinear
contributions are contained in the factor $\sqrt{Z_i(m)}$ that is
process independent and known~\cite{Engel:2018fsb} and has to be added
for each external electron. Finally, ultrasoft contributions exist for
individual integrals and diagrams, but they cancel for the amplitude
in agreement with the SCET expectations.
{\hbadness=10000

}
This result can now be combined with a fully massive evaluation of
$d\sigma^{(\mathrm{rv})}(m)$ and $d\sigma^{(\mathrm{rr})}(m)$.
However, the fully massive real corrections contain poles that will
not naively match the poles obtained through massification, instead
causing a mismatch at $\mathcal{O}(z)$. This mismatch can be avoided
by either expanding the analytic poles of the real corrections or
calculating the fully massive poles of the two-loop amplitude from
first principles~\cite{Yennie:1961ad,Becher:2009kw}. For the
phase-space integration only soft singularities have to be regularised
with dimensional regularisation.  Putting everything together results
in a Monte Carlo code that provides results complete at NLO and
includes all leading in $z$ terms at NNLO. However, it does not
systematically include non-leading terms at NNLO, i.e. terms that
vanish in the limit $z\to 0$, such as $\alpha^2 z \log(z)$.  It should
also be mentioned that in the region $t\to -0$ the counting used in
this expansion breaks down. Of course the cross section is divergent
in this region anyhow such that this problem can be avoided with an
appropriate cut.

\subsection{Beyond next-to-next-to-leading order}
\label{sec:Bnnlo}

A complete calculation at N$^3$LO is a daunting task. However, keeping
in mind that the dominant contribution to any loop order stem from
emission of the electron, i.e. terms $\mathcal{O}(q^{2+2n}Q^2)$, at
least a partial N$^3$LO result might be achievable. Once more it is
the remarkable simplicity of QED that allows us to extend the
subtraction scheme proposed for NNLO, FKS$^2$, to even higher loop
orders~\cite{Engel:2019nfw}. The N$^3$LO extension, aptly named
FKS$^3$, has already been worked out and shown to retain the
simplicity of FKS.

The necessary ingredients for this endeavour are the $\A713$ part of
the three-loop $\Amp{n}{3}$, the $\A612$ part of the two-loop
$\Amp{n+1}{2}$, the $\A511$ part of the one-loop $\Amp{n+2}{1}$, and
the $\A411$ part of the tree-level $\Amp{n+3}{0}$. The latter two are,
at least in principle, easy to obtain thanks to the advances made in
the automation of one-loop calculations. The former two are more
challenging. However, impressive progress has been made in calculating
the heavy-quark form factors at three-loop~\cite{Ablinger:2018yae,
  Blumlein:2019oas} and an efficient tool to numerically evaluate
generalised polylogarithms~\cite{Naterop:2019xaf} is available. A big
remaining problem is the two-loop real-double-virtual contribution.
However, at least in principle, it should be possible to adapt and
massify the calculations performed for $\gamma^*\to qqg$ which are
part of the NNLO calculations to three-jet production.\par

\section{Hadronic Vacuum Polarisation Contributions} \label{hvp}

\subsection{Next-to-leading order}\label{hvp:nlo}

\begin{figure}
\centering
\subfloat[NLO/FSR \label{fig:hadloFSR}]{
    \includegraphics[width=0.3\columnwidth]{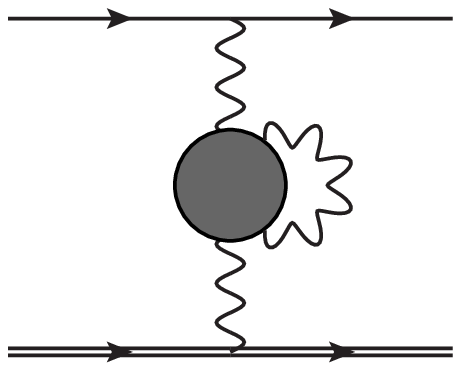}}\quad
    \subfloat[class I\label{fig:classI}]{
    \includegraphics[width=0.3\columnwidth]{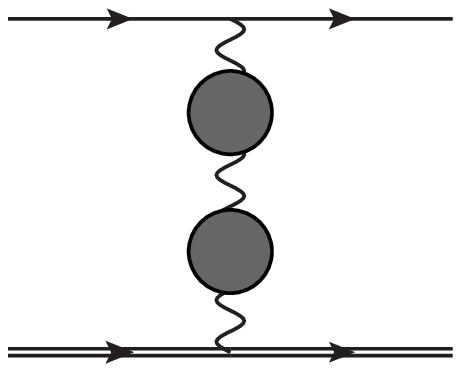}}\quad
  \subfloat[class II\label{fig:classII}]{
    \includegraphics[width=0.3\columnwidth]{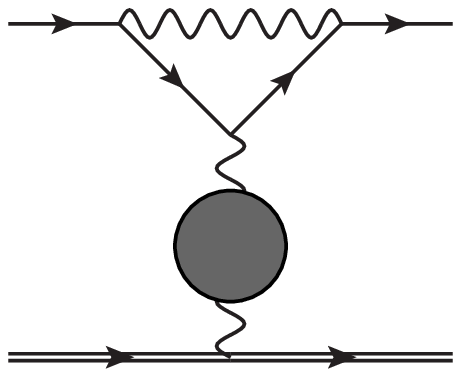}} \\
  \subfloat[class III\label{fig:classIII}]{
    \includegraphics[width=0.3\columnwidth]{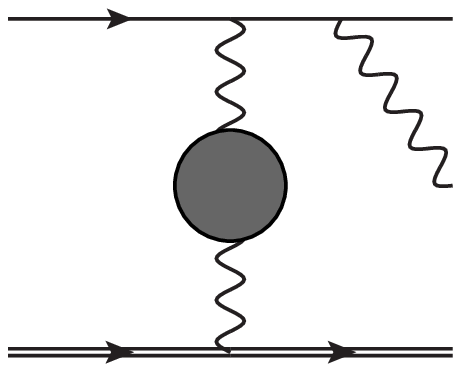}} \quad
  \subfloat[class IV\label{fig:classIV}]{
  \includegraphics[width=0.3\columnwidth]{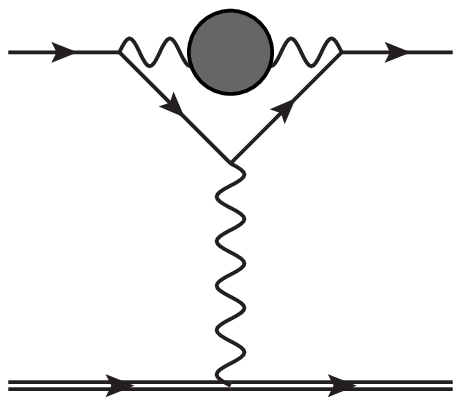}\quad
  \includegraphics[width=0.3\columnwidth]{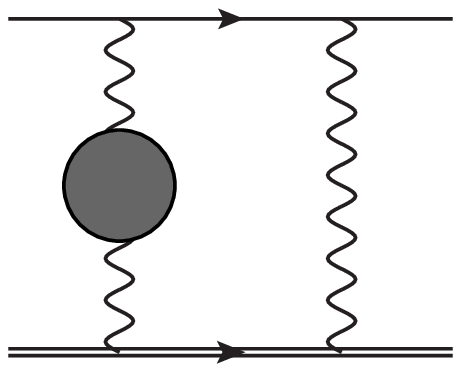}}
  \caption{(a) Diagram contributing to the hadronic correction to
    $\mu$-$e$ scattering at NLO. (b--e) Examples of diagrams
    contributing to the four classes of hadronic corrections at
    NNLO. Electrons, muons and photons are depicted with thin, thick
    and wavy lines, respectively. The grey blobs indicate hadronic
    vacuum polarisation insertions.}
  \label{fig:fd}
\end{figure}

The NLO and NNLO corrections to the muon-electron differential cross
section involve non-perturbative QCD contributions given by diagrams
with an HVP insertion in the photon propagator (see
Figs.~\ref{fig:lo:hvp} and~\ref{fig:fd}). Let us define the SM vacuum
polarisation tensor with four-momentum $q$ as
\begin{align}
  i \Pi^{\mu\nu}(q) &= i \Pi(q^2) (g^{\mu\nu} q^2-q^\mu q^\nu)
  \nonumber\\
  &=  \int d^4 x \, e^{iqx} \bra{0}
  T\{j_\mathrm{em}^\mu(x) j_\mathrm{em}^\nu(0) \} \ket{0},
\end{align}
where $j_\mathrm{em}^\mu(x) = \sum_f Q_f \bar{\psi}(x)\gamma^\mu
\psi(x)$ is the electromagnetic current and the sum runs over fermions
with charges $Q_f$.  The function $\Pi(q^2)$ is the renormalised
vacuum polarisation satisfying $\Pi(0) = 0$. It is commonly subdivided
into the leptonic part $\Pi_\lep$, which receives contributions only
from charged leptons, the hadronic part $\Pi_\had$, from hadrons
containing the five light quarks $u,d,s,c,b$, and the contribution
from the top quark $\Pi_\mathrm{top}$. The weak interaction will be
ignored.

In perturbation theory $\Pi_\lep$ and $\Pi_\mathrm{top}$ can be
computed order by order in $\alpha$ and the strong coupling
$\alpha_s$~\cite{Broadhurst:1993mw, Kuhn:1998ze, Steinhauser:1998rq,
  Sturm:2013uka}. On the contrary, the HVP cannot be calculated in
perturbation theory for any value of $q^2$ because of the
non-perturbative nature of strong interactions. Nevertheless, we can
express $\Pi_\had$ in terms of the measured cross section of the
reaction $e^+e^- \to$~hadrons~\cite{Jegerlehner:2017gek} thanks to the
subtracted dispersion relation and the optical theorem
\begin{align}
  \frac{\Pi_\had(q^2)}{q^2} =
  -\frac{\alpha}{3 \pi}
  \int_{4m_\pi^2}^\infty
  \frac{dz}{z}
  \frac{R(z+i \varepsilon)}{q^2-z+i\varepsilon},
  \label{eqn:disprel}
\end{align}
where
\begin{align}
  R(s) = \frac{\sigma(e^+e^- \to \mathrm{hadrons})}{
    4\pi|\alpha(s)|^2/(3s)}
\end{align}
and
\begin{align}
  \alpha(s) = \frac{\alpha}{1-\Delta \alpha(s)}
\end{align}
is the effective fine-structure constant.  The numerical value for the
HVP can be obtained by using the Fortran libraries
\texttt{alphaQEDc19} \cite{Jegerlehner:2001ca, Jegerlehner:2006ju,
  Jegerlehner:2011mw, Harlander:2002ur},
\texttt{KNT18VP}~\cite{Hagiwara:2003da, Hagiwara:2006jt,
  Hagiwara:2011af, Keshavarzi:2018mgv, Harlander:2002ur,
  Actis:2010gg}, as well as \texttt{VPLITE}~\cite{Actis:2010gg, Ignatov:2016}
based on hadronic $e^+ e^-$ annihilation (timelike) data.

The hadronic contribution to the $\mu$-$e$ cross section at NLO, due to
the diagram in Figure~\ref{fig:lo:hvp}, is
\begin{equation}
  \frac{d \sigma^\mathrm{NLO}_\had}{dt} = 
  -2 \, \Pi_\had(t) \, \frac{d\sigma^{(0)}}{dt} = 
    2 \, \Delta \alpha_\had(t) \, \frac{d\sigma^{(0)}}{dt},
  \label{eqn:xsechadlo}
\end{equation}
where $\Delta \alpha_\had(t) = - \Pi_\had(t)$ is the leading hadronic
contribution to the running of $\alpha(t)$. The goal of the MUonE
experiment is the extraction of $\Delta \alpha_\had(t)$ from $\mu$-$e$
scattering data.  Note that the NLO hadronic corrections incorporate
also the contribution from the diagram in Fig.~\ref{fig:hadloFSR}
where a virtual photon is emitted and reabsorbed by the hadronic
insertion. This irreducible part of the second-order hadronic
contribution to the running of $\alpha(t)$ is not considered as part
of the NNLO corrections because its effect is commonly included in the
ratio $R(s)$ as final-state radiation~\cite{Melnikov:2001uw,Passera:2004bj}.

\begin{figure}
  \centering
  \includegraphics[width=0.45\textwidth]{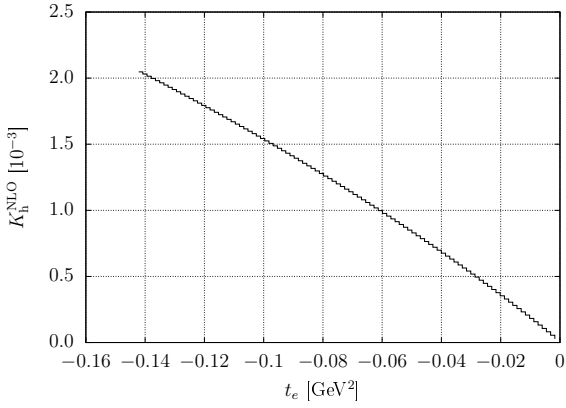}
  \caption{The relative importance of the HVP at NLO in $\mu$-$e$
    scattering as a function of $t_e$.}
  \label{fig:hvpnloB}
\end{figure}

The impact of the hadronic contribution at NLO is shown in
Figure~\ref{fig:hvpnlo} as a function of $\theta_e$. For later
reference, in Figure~\ref{fig:hvpnloB} we show the same contribution
as a function of $t_e$. The factor $K_\had^{\mathrm{NLO}}(t_e)$
depicted in Figure~\ref{fig:hvpnloB} is defined in analogy to
\eqref{eqn:RhoHNLO}.  In accordance with Figure~\ref{fig:hvpnlo}, the
effect is larger for large values of $|t_e|$.

Before we move on to the hadronic corrections at NNLO, following the
analysis of~\cite{Masiero:2020vxk} we will briefly discuss the impact
of the SM corrections -- and possibly NP -- on the extraction of
$\Delta \alpha_\had(t)$ at MUonE. This experiment will extract $\Delta
\alpha_{\had} (t)$ from the shape of the differential $\mu$-$e$
scattering cross section by a template fit method~\cite{LoI}. The
basic idea is that $\Delta \alpha_\had (t)$ can be obtained measuring,
bin by bin, the ratio $N_i/N_{\rm {n}}$, where $N_i$ is the number of
scattering events in a specific $t$-bin, labelled by the index $i$, and
$N_{\rm {n}}$ is the number of events in the normalization $t$-bin
corresponding to $x(t) \sim 0.3$ (for this value of $x$, $\Delta
\alpha_\had (t)$ is comparable to the experimental sensitivity
expected at MUonE and its error is negligible). Therefore, this
measurement will not rely on the absolute knowledge of the
luminosity. To extract the leading hadronic corrections to the
$\mu$-$e$ scattering cross section in the $t$-bin $i$, one can split
the theoretical prediction into
\begin{align}
  \sigma_{{\rm {th}},i} = \sigma^{(0)}_{i} \Big[ 1 + 2 \Delta
    \alpha_{{\had},i}
    + \delta_{i} + \delta_{{\rm NP},i} \, \Big],
\end{align}
where $\sigma^{(0)}_{i} = \int_i (d \sigma^{(0)}/ dt ) dt$ is the LO
QED prediction integrated in the $t$-bin $i$, $2 \Delta
\alpha_{{\had},i}$ is the leading hadronic correction obtained
from~\eqref{eqn:xsechadlo}, $\delta_i$ is the remainder of the SM
corrections, and $\delta_{{\rm NP},i}$ is a possible NP
contribution. The experimentally measured ratio $N_i/N_{\rm {n}}$ can
then be equated with the ratio of the theoretical predictions,
\begin{align}
 \frac{N_i}{N_{\rm {n}}}  = 
 \frac{\sigma_{{\rm th},i}}{\sigma_{\rm {th,n}}}
	\simeq
  \frac{\sigma^{(0)}_i}{\sigma^{(0)}_{\rm {n}}} \, \Big[ 
   1 &+ 2 \left( \Delta \alpha_{{\rm had},i} - \Delta
   \alpha_{{\rm had,n}}\right)    \nonumber \\
   &+ \left(\delta_i -\delta_{\rm n} \right)+ 
     \left(\delta_{{\rm NP},i} - \delta_{\rm NP,n} \right ) \Big] .
\label{eqn:norm}
\end{align}
As $\Delta \alpha_{{\rm had, n}}$ is known with negligible error, if
$\left(\delta_i -\delta_{n} \right)$ is computed with sufficient
precision, one can extract
\begin{align}
  2 \Delta \alpha_{{\had},i} +
  \left(\delta_{{\rm NP},i}-\delta_{\rm NP,n} \right )\, ,
  \nonumber
\end{align}
bin by bin,
from $N_i/N_{\rm {n}}$. Equation~(\ref{eqn:norm}) shows that the
impact of the SM corrections on this extraction can only be
established after subtracting their value in the normalization region,
and that the MUonE experiment will not be sensitive to a NP signal
constant in $t$ relative to the LO QED one, i.e.\ such that
$\delta_{{\rm NP},i} = \delta_{\rm NP,n}$~\cite{Masiero:2020vxk}.

\subsection{Next-to-next-to leading order}

At NNLO, we split the hadron-induced corrections to $\mu$-$e$
scattering, of order $\alpha^4$, into four classes of diagrams.  The
first three classes contain \emph{factorisable} contributions,
i.e.\ amplitudes that can be written as the product of a QED amplitude
times the function $\Pi_\had(q^2)$ evaluated at some $q^2$ fixed by
the external kinematics. They are:
{\hbadness=10000
\begin{description}
\item[Class I:] tree-level diagrams in combination with one or
  two vacuum-polarisation insertions (Figure~\ref{fig:classI}). Their
  contribution to the differential cross section is proportional to
  $\Pi_\had(t)[\Pi_\had(t)+2\Pi_\lep(t)]$, the reducible part of the
  se\-cond-order hadronic contribution to the running of $\alpha(t)$.
\item[Class II:] QED one-loop diagrams in combination with one HVP
  insertion in the $t$-channel photon
  (Figure~\ref{fig:classII}). Their contribution to the differential
  cross section is proportional to $\Pi_\had(t)$ and a combination of
  one-loop QED corrections to $\mu$-$e$ scattering.
\item[Class III:] real photon emission diagrams with a
  vacuum-polarisation insertion in the $t$-channel photon
  (Figure~\ref{fig:classIII}). They contain terms proportional either
  to $\Pi_\had(t_e)$ or to $\Pi_\had(t_\mu)$.
\end{description}}
Moreover a fourth class of \emph{non-factorisable} diagrams must be
considered:
{\hbadness=10000
\begin{description}
\item[Class IV:] one-loop QED amplitudes with a hadronic vacuum
  polarisation insertion in the loop. They can be further subdivided
  into vertex and box corrections (Figure~\ref{fig:classIV}).
\end{description}
}%

There are no light-by-light contributions to the $\mu e$ cross section
at NNLO (order $\alpha^4$) -- they appear at N${}^{3}$LO (order
$\alpha^5$).
In addition, the analysis of future $\mu$-$e$ scattering data will also
require the study of $\mu$-$e$ scattering processes with final states
containing hadrons, as for instance $e^-\mu^\pm \to e^- \mu^\pm
\pi^+\pi^-$ and $e^- \mu^\pm \to e^- \mu^\pm \pi^0$. However, as
$\sqrt{s}\simeq 400$ MeV, the available phase space is quite small:
$\sqrt{s}-M-m-2m_{\pi^0}\simeq 20$ MeV for the former and
$\sqrt{s}-M-m-m_{\pi^0}\simeq 160$
MeV for the latter process.

As the HVP per se is of non-perturbative nature, the hadronic
NNLO corrections rely inevitably on some external data for their
numerical evaluation.  These inputs can be of two kinds: we can either
use the $R$ ratio and the traditional dispersive method, or we can
dismiss the $e^+e^- \to$ hadron data --- after all, MUonE aims at
measuring $a_\mu^\mathrm{HLO}$ independently on $R$ --- and employ the
very same space-like data measured by MUonE.  The two approaches are
the following.
\begin{description}[leftmargin=14pt]
  \item[\boldmath To $R$:]
The traditional approach to calculate the amplitudes in class IV uses
the dispersion relation to replace the dressed photon propagator
inside the loop --- where $q$ now stands for the loop momentum ---
with the r.h.s.\ of \Eqn{eqn:disprel}, where the momentum $q$
appears only in the term $1/(q^2-z)$. Therefore, the dispersion
relation effectively replaces the dressed propagator with a massive
one, where $z$ plays the role of a fictitious squared photon mass.
This allows to interchange the integration order and evaluate, as a
first step, the one-loop amplitudes with a ``massive'' photon. The
results obtained for the $z$-dependent scattering amplitudes are then
convoluted with the $R$ ratio.  Also for the amplitudes in classes
I-III we rely on the dispersion relation~\Eqn{eqn:disprel} to compute
the HVP in the space-like region.  This method was employed, for
example, to compute the hadronic corrections to muon
decay~\cite{vanRitbergen:1998hn,Davydychev:2000ee} and Bhabha
scattering~\cite{Actis:2007fs,Kuhn:2008zs,CarloniCalame:2011zq}.

The hadronic NNLO corrections to $\mu$-$e$ scattering based on the $R$
ratio were presented in~\cite{Fael:2019nsf}. Two independent codes
were developed. The first is a standard Monte Carlo which uses
\texttt{Collier}~\cite{Denner:2016kdg} for the evaluation of the
one-loop tensor integrals and employs the FKS subtraction
scheme~\cite{Frixione:1995ms,Frederix:2009yq}. The second code is
developed in \textsc{Mathematica} and takes advantage of the analytic
expressions of the one-loop integrals from
\texttt{Package-X}~\cite{Patel:2015tea} and \textsc{Mathematica}'s
arbitrary-precision numbers to check for numerical instabilities
during the dispersive and phase-space integrations. Perfect agreement
was found between the two implementations. The results are presented
below.

\item[\boldmath Not to $R$:]
This alternative approach was presented in~\cite{Fael:2018dmz}. The
factorisable diagrams in classes I, II and III depend on
$\Pi_\had(t)$, which is the quantity extracted by MUonE from the
diagram in Figure~\ref{fig:lo:hvp}. As discussed
in~\cite{Fael:2018dmz}, also the non-factorisable corrections in class
IV, where the vacuum polarisation appears inside a loop, can be
calculated employing the HVP in the space-like region, without making
use of the $R$ ratio.

Indeed, the loop integrals containing $\Pi_\had$ can be computed via
the hyperspherical integration method.  After introducing spherical
coordinates for the loop momentum and continuing internal and external
momenta to the Euclidean region, one can write the loop propagators as
an expansion in Gegenbauer polynomials.  Then, the integration over
the angular variables is performed analytically thanks to the
orthogonality properties of these polynomials, so that each diagram is
eventually cast in the form of a residual radial integration,
\begin{equation}
  \int_0^\infty dQ^2 \, Q^2 \, \Pi_\had(-Q^2) \, f(Q^2,s,t),
\end{equation}
which is computed numerically once provided with the HVP in the
space-like region. The expressions of the kernels $f(Q^2,s,t)$ were
presented in~\cite{Fael:2018dmz}. Their implementation into a
numerically stable code is necessary for future use in the Monte
Carlo.

The fact that the hadronic NNLO corrections can be obtained from
$\Pi_\had(q^2)$, with just $q^2<0$, suggests the possibility for MUonE
to determine the HVP in an iterative way without making use of the $R$
ratio. As a first step the hadronic NNLO corrections can be switched
off in the Monte Carlo and a first approximation for $ \Pi_\had(q^2)$
extracted. Afterwards, the Monte Carlo can be supplied with such first
approximation to compute the hadronic NNLO corrections, then a second
approximation ex\-trac\-ted and the process further iterated.

Alternatively, if a functional form for $\Pi_\had(q^2)$ is chosen to
fit the HVP~\cite{pagani2017} the same ansatz can be employed at NNLO,
under the assumption that it satisfies the correct asymptotic behaviour
at infinity.

\end{description}
\bigskip

The dispersive and the hyperspherical methods are of course identical
from the mathematical point of view; however the \emph{to $R$} and the
\emph{not to $R$} approaches differ for the underlying theoretical
assumptions.  If we use the $R$ ratio, we make a distinction between
the HVP entering at NLO and at NNLO. On the one hand, at NLO we leave
the HVP in a free form to be fitted from data. On the other hand, at
NNLO we choose a different $\Pi_\had(q^2)$ whose values are given by
the $R$ ratio via the dispersion relation.

On the contrary, by employing the hyperspherical method in the
\emph{not to $R$} approach we treat the HVP in a consistent way to all
orders without making any a priori assumptions. Moreover, only in
latter case, the MUonE determination of $a_\mu^\mathrm{HLO}$ becomes
truly independent and completely uncorrelated from time-like
measurements.

\begin{figure}
\centering
    \includegraphics[width=0.45\textwidth]{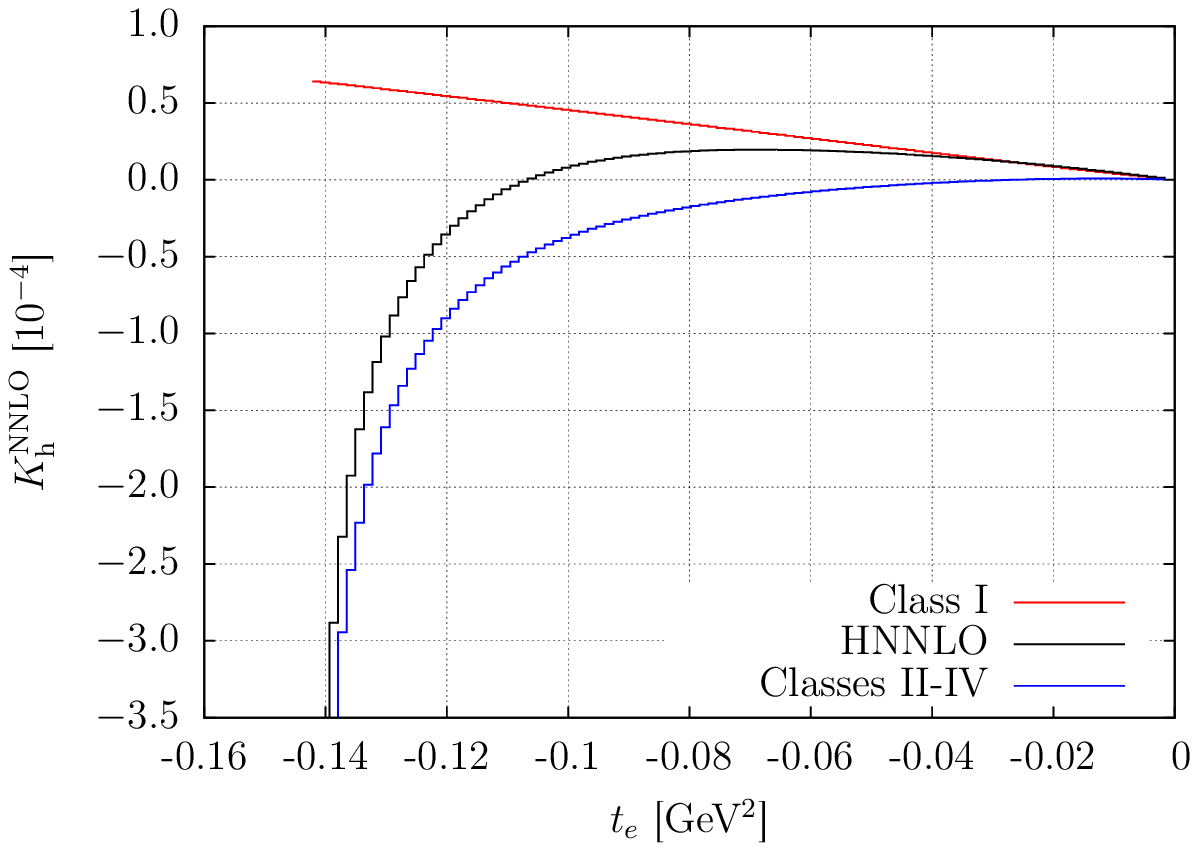}\quad
    \includegraphics[width=0.45\textwidth]{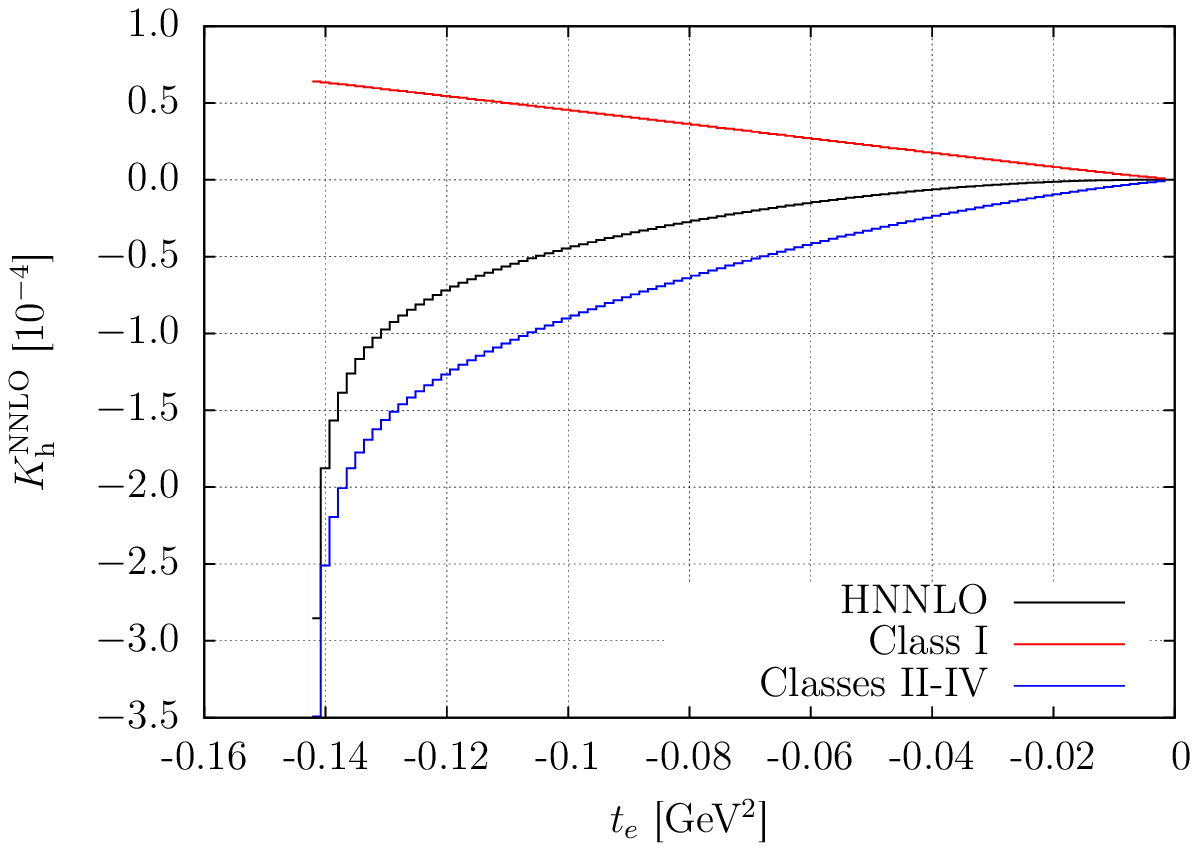}
  \caption{$K_\had^{\mathrm{NNLO}}(t_e)$ factor for a positive (upper
    panel) and negative (lower panel) muon beam of energy
    $E_1=150$~GeV. The total hadronic NNLO correction are depicted in
    black, while the contributions of class I (II-IV) are shown
    separately in red (blue).}
  \label{fig:K}
\end{figure}
Let us now discuss the size of these hadronic corrections. The ratio
of the NNLO hadronic contribution to the $\mu e$ differential cross
section, with respect to the squared momentum transfer $t_e$, and the
LO prediction,
\begin{align}
	K_\had^{\mathrm{NNLO}}(t_e) = 
 \frac{d \sigma_\had^{\mathrm{NNLO}}}{dt_e} / \frac{d \sigma^{(0)}}{dt_e},
\label{eqn:RhoHNNLO}
\end{align}
is shown in Figure~\ref{fig:K} for the processes $\mu^+ e^- \to \mu^+
e^-$ (upper panel) and $\mu^- e^- \to \mu^- e^-$ (lower panel), where
we use $E_1=150$~GeV.
The corrections shown in Figure~\ref{fig:K} were computed
in~\cite{Fael:2019nsf} using the dispersive approach and employing
\texttt{alphaQEDc17} for the numerical evaluation of the HVP.  The
black lines indicate the total hadronic contribution arising from
classes I--IV, while the blue ones show the sum of the contributions
of classes II, III, and IV, but not I.
Figure~\ref{fig:K} shows that for a muon beam with energy 150~GeV,
most of the kinematic region scanned by the momentum transfer $t_e$
results in a factor $K_\had^{\mathrm{NNLO}}(t_e)$ which is of order
$10^{-4}$--$10^{-5}$. These corrections are therefore larger than the
$O(10^{-5})$ precision expected at the MUonE experiment. The ratio
$K_\had^{\mathrm{NNLO}}(t_e)$ contains a term that diverges
logarithmically at the end of the electron spectrum. This feature,
clearly visible in Figure~\ref{fig:K} for $t_e \to t_{\rm min}$, is
related to the infrared divergence, indicating a breakdown of the
perturbative expansion and the need for resummation.

The uncertainty on $K_\had^{\mathrm{NNLO}}$ due to the error on $R$
was estimated by comparing the values obtained with the libraries
\texttt{alphaQEDc17} and \texttt{KNT18VP}. For each value of $t_e$, we
found that the relative difference between the two calculations of
$d\sigma_\had^\mathrm{NNLO}/dt_e$ is about 1\% or less. Therefore a
relative uncertainty of 1\% was assigned on
$d\sigma_\had^\mathrm{NNLO}/dt_e$, which corresponds to an error in
$K^\mathrm{NNLO}_\had (t_e)$ of $O(10^{-6})$ or less, well below the
precision expected at the MUonE experiment.

\section{Beyond fixed order} \label{resummation}

In Section~\ref{fixedorder} we have discussed the computation of the
cross section in a strict expansion in the coupling $\alpha$. At
N$^n$LO the cross section contains large logarithms of the form
\begin{align}
\label{def:Lm}
\alpha^2\times (\alpha L_m)^n \equiv
\alpha^2\times\alpha^n \log^n\frac{m^2}{Q^2}
\end{align}
that potentially invalidate the perturbative expansion since $\alpha
L_m$ is not necessarily a good expansion parameter.  The prefactor
$\alpha^2$ is from the Born cross section. In the total cross
section the only logarithms that survive are due to initial-state
collinear emission from the electron, as discussed in
Section~\ref{fixedorder}. However, for differential cross sections
final-state collinear logarithms $L_m$ can be present.

Going beyond the total cross section, as is required for MUonE, can
result in additional large logarithms. In order to select elastic
scattering, the emission of real radiation has to be restricted.
Naively, this is done by vetoing photon emission with energy larger
than a cutoff $\Delta$. For the moment we ignore the fact that in
practice another observable has to be chosen since photons are not
detected.  Restricting the emission of real radiation will result in
additional logarithms of the form $L_\Delta = \log(\Delta^2/Q^2)$
which again can be large if the cut is severe, i.e. $\Delta$ is
small. Thus for each order in $\alpha$ we obtain up to two powers of
large logarithms. Of course, this is closely related to the
$1/\epsilon^2$ (or the $ 1/\epsilon\, \log(m)$) singularities for each
perturbative order, discussed in Section~\ref{fixedorder}. Therefore,
{\hbadness=10000
the $n$th order correction to a differential cross section has the
structure
\begin{align}
\label{sigmalogs}
d\sigma^{(n)} = \alpha^2\times
 \Big(\frac{\alpha}{\pi}\Big)^n\, \sum_{n_1,n_2} c^{(n)}_{n_1,n_2} \,
\big(L_m\big)^{n_1} \big(L_\Delta\big)^{n_2} \, ,
\end{align}
}
where the sum runs over $0\le n_1, n_2 \le n$. Again, the prefactor
$\alpha^2$ is due to the Born term. In what follows we will omit this
factor when discussing powers of couplings and logarithms.

Another potential source of large logarithms is related to the
so-called factorisation (or collinear) anomaly~\cite{Beneke:2005,
  Chiu:2011qc, Becher:2011pf}. This is related to the breaking of a
scaling symmetry between collinear and soft modes in SCET and occurs
due to the presence of two non-vanishing
masses~\cite{Engel:2018fsb}. In practice it means that the separate
factors of \eqref{factorizemM} might contain singularities that are not
regularised through the usual dimensional regularisation. While these
singularities cancel between the various factors on the r.h.s. of
\eqref{factorizemM} the left-over of these cancellations corresponds
to a logarithm of the form $\log(m M/Q^2)$.

\subsection{Leading logarithm}

The terms of~\eqref{sigmalogs} with $n_1=n_2=n$ are the leading
logarithms (LL). They can be resummed using a parton shower (PS). A PS
has the advantage that the kinematics of the emitted photons is
retained so that exclusive events can be generated. This makes sure
that the resulting program remains fully-differential in all resolved
particles which cannot be guaranteed in analytic calculations.

Roughly speaking, there are two avenues to numerically resum the LL
contributions. The starting point is either soft emission or collinear
emission. In the first case, the well-known Yennie Frautschi Suura
(YFS) exponentiation~\cite{Yennie:1961ad} of soft emission is
used. This allows for a numerical implementation taking into account
soft emission to all orders~\cite{Barberio:1993qi, Jadach:1995nk,
  Hamilton:2006xz, Schonherr:2008av}. Such a resummation is well
suited to be combined with fixed-order calculations performed with
FKS$^2$, the subtraction scheme suggested earlier. In fact, FKS$^2$
exploits the YFS structure of the matrix elements
\begin{align}
\label{eq:yfs}
\sum_{\ell = 0}^\infty \M{n}{\ell} = 
e^{-\alpha \hat{\mathcal{E}}}\, \sum_{\ell = 0}^\infty
\mathcal{M}_{n}^{f(\ell)} \,
\end{align}
where $\mathcal{M}_{n}^{f(\ell)}$ is free of IR singularities. The
latter are all absorbed by the exponential of the integrated eikonal
$\hat{\mathcal{E}}$ that governs soft emission. For a precise
definition of all quantities in \eqref{eq:yfs} and more details
see~\cite{Engel:2019nfw}. A recent example where a NNLO QED
calculation is merged with a YFS resummation can be found e.g.
in~\cite{Krauss:2018djz}.

Taking collinear emission as a starting point, a QED parton shower can
be constructed through subsequent collinear emission of photons
governed by the $e\to e\,\gamma$ splitting kernel
$P(z)=(1+z^2)/(1-z)$, where $z$ is the momentum fraction of the
electron after the split. This procedure has been used by the
BabaYaga~\cite{Balossini:2006wc, Balossini:2008xr,
  CarloniCalame:2006zq, CarloniCalame:2007cd, Boselli:2015aha} event
generator. It can be combined with fixed-order calculations and
extended to next-to-leading collinear logarithms.

As both, YFS Monte Carlo and QED parton shower include the leading
soft-collinear emissions, they agree at LL. Going beyond LL, the QED
PS also includes hard (non soft) collinear radiation, i.e. it includes
all leading collinear logs $\alpha^n L_m^n$. It can be further adapted
to also include soft wide-angle emission~\cite{CarloniCalame:2001ny}.
As we will discuss below and in Section~\ref{montecarlo}, this
difference beyond various implementations at LL can reveal useful
information to assess the theoretical error. To exploit this, work is
ongoing to implement $\mu$-$e$ scattering in both frameworks and
compare.

\subsection{Next-to-leading logarithm}

The terms of~\eqref{sigmalogs} with $n_1+n_2 = 2 n-1$ are the
next-to-leading logarithms (NLL). Whether or not their resummation is
required and possible depends on the precise definition of the
quantity that selects elastic scattering and on the value of the cut
parameter $\Delta$. A partial resummation of NLL terms can be done
with improved Monte Carlo generators.

A complete resummation beyond LL requires a precise definition of the
physical quantity for which the resummation is carried out. Given that
for $\mu$-$e$ scattering we are not primarily interested in particular
distributions of certain observables, but rather in a precise
description of the fiducial cross section measured by MUonE, it is not
possible to precisely match the observable to the measured
quantity.

The most important cut that has to be made for the extraction of the
HVP is to choose elastic events. In theory this can be achieved in
several ways, all of which restrict the phase space for emission of
photons. As a first example we mention a cut on the invariant mass
$m_{e\gamma}$ of the electron-jet, i.e. the cluster of the outgoing
electron plus all potentially emitted photons. This quantity is
sensitive to large-angle soft emission and, contrary to
e.g. $m^2_{\mu\gamma}$, also to small-angle hard emission. We thus
define
\begin{align}
  \label{meinv}
  m^2_{e\gamma} = (p_4+p_X)^2 = (p_1+p_2-p_3)^2\, ,
\end{align}
where according to \Eqn{process} $p_X = \sum_{i=1,n} p_{i\gamma}$ is
the sum over (up to $n$) photon momenta. The elastic events can be
chosen by making a cut $m_{e\gamma} - m \le \Delta$.  Another
well-studied option is to use the transverse momentum. In our case we
have to take the transverse momentum $p_{4\perp}$ of the electron
w.r.t. its tree-level direction, which can be determined from the muon
scattering angle $\theta_\mu$. In case of no photon emission,
$p_{4\perp} \to 0$ and we can impose a cut $|p_{4\perp}|^2 < \Delta^2$
to select elastic events.

While these quantities are useful from a theoretical point of view and
likely enable a resummation beyond LL, they are unfortunately not very
useful from an experimental point of view. None of these quantities
can actually be measured since neither are photons detected nor are
the momenta or energies of the electron and muon measured.  In
practice, the experimental procedure to select elastic events has to
rely solely on the scattering angles.  Such a quantity is likely to be
rather involved and not amenable to direct resummation. In
Section~\ref{montecarlo} we will consider an acoplanarity cut as one
example of a cut that can realistically be applied by MUonE to
restrict radiative events.

A possible way forward is to use the analytic resummation of several
different variables to construct approximations for perturbative
coefficients beyond those included in the fixed-order approach. These
results can then be implemented as approximate matrix elements in a
fully differential parton-level Monte Carlo. Through comparisons of
results obtained by using different versions of resummation it is
possible to obtain a realistic error estimate of the
approximations. This procedure has been used for example for top-quark
pair production~\cite{Broggio:2014yca} and provided an improved
prediction with a robust estimate of missing terms. A quantity that
offers itself for resummation in the context of $\mu$-$e$ scattering
is $d\sigma/dt_e$ where in the region $t_e\to t_\mathrm{min}$ large
logarithmic corrections are present.

\subsection{Estimate of the theory error}

As discussed earlier in the present section, higher-order
contributions can be included according to different methods, such as
a QED PS algorithm, YFS MC exponentiation or analytic
resummation. Regardless of the approach used to account for the
corrections beyond NNLO, the accuracy of the theoretical predictions
due to missing perturbative contributions must be carefully estimated,
as it represents a component of the total systematic error.

For this purpose, it seems advisable to evaluate the theoretical
uncertainty step-by-step, as the different theoretical ingredients
become available. In the following, a possible strategy for the
theoretical uncertainty estimate is illustrated, at the level of
differential distributions. It is assumed that, in addition to the
fixed-order NLO QED calculation, also the NNLO QED matrix elements are
implemented in a fully fledged Monte Carlo simulation tool. The
technical accuracy, related to the details of the implementation of
the fixed-order radiative corrections, can be controlled by means of
two completely independent codes, which are assumed to exist.

In the relatively short term, a first assessment of the theoretical
accuracy could be given as follows:
\begin{itemize}

\item by comparing the predictions for the photonic corrections at NLO
  and NNLO accuracy. This comparison can be performed for the full set
  of corrections but also separately for the gauge-invariant subsets
  of contributions due to electron radiation, muon radiation and
  electron-muon interference. The importance of this procedure is
  twofold as {\it a)} it would allow to settle the hierarchy of the
  different classes, which is a crucial prerequisite to identify the
  sources of corrections that need to be resummed at all orders and
  {\it b)} it would provide information about the convergence of the
  perturbative series (in particular, for those kinematical regions
  where the NLO corrections are particularly large). A first naive
  estimate of the missing third order can be given by
\begin{equation}
  \frac{d\sigma^{\rm N3LO} - d\sigma^{\rm NNLO}}{d\sigma^{\rm NNLO} - d\sigma^{\rm NLO}} \simeq 
  \frac{d\sigma^{\rm NNLO} - d\sigma^{\rm NLO}}{d\sigma^{\rm NLO} - d\sigma^{\rm LO}}.
\end{equation}
\item by computing the NNLO leptonic and hadronic corrections due to
  the combination of the two-loop vacuum polarisation contribution and
  real pair emission. It is known that these corrections give rise to
  large collinear logarithms but also that they are typically smaller
  than purely photonic corrections. In particular, the contribution
  due to electron loop and real $e^+ e^-$ radiation produce collinear
  logarithms $L_m$. Taken separately, the LO cross section of the
  process $e\mu\to e\mu (e^+ e^-)$ results in contributions $\alpha^2
  L_m^3$. However, if combined with the virtual electron-loop
  contributions the $\alpha^2 L_m^3$ cancel and we are left with
  collinear logarithms $\alpha^2 L_m^2$ and $\alpha^2 L_m$.  The
  computation of this class of corrections therefore would allow to
  probe the size of those NNLO logarithmically-enhanced corrections
  that are of non-photonic nature;
\item by comparing the finite-order expansion of a given resummation
  approach with the exact perturbative calculation at NNLO
  accuracy. Again, this comparison could be performed for the complete
  set or the gauge-invariant subsets of photonic corrections and would
  allow to quantify the size of the NNLO remainder beyond the LL
  approximation at ${\cal O} (\alpha^2)$.

\end{itemize}

Over the longer term, assuming that different methods to account for
the contribution of multiple photon radiation will be available and
matched to the NNLO calculation, the theoretical uncertainty could be
more reliably estimated as follows
{\hbadness=10000

\begin{itemize}

\item through a comparison of the exact NNLO calculation and the
  ${\cal O} (\alpha^3)$ expansion of a given resummation
  procedure. From this comparison, one would get an evaluation of the
  whole set of higher-order contributions beyond NNLO;

\item by comparing the all-order predictions of the different methods
  developed for the description of multiple photon radiation. As
  remarked earlier, approaches such the QED PS and YFS MC
  exponentiation provide the same LL structure but may differ in the
  partial resummation of NLL contributions. Moreover, this comparison
  could be extended, wherever possible, to include the results of
  analytic resummations possibly featuring a complete resummation
  beyond the LL approximation. As a whole, this procedure would
  provide a robust estimate of missing higher-order terms at NLL
  accuracy;

\item under the assumption that two independent implementations of a
  MC code based on the matching of NNLO corrections with resummation
  will be available and different techniques for exponentiation will
  be used, a comparison between the predictions of the two codes could
  provide further important information about the theoretical
  accuracy.  Actually, because of the reasons already emphasised, the
  two calculations are expected to differ for contributions dominated
  by terms of the order of $\alpha^3 L_m^2$. Hence, this comparison
  would allow to probe the size of the most important NLL ${\cal O}
  (\alpha^3)$ contributions, similarly to the previous point above but
  at the level of completely matched formulations.

\end{itemize}
}

Besides the contributions due to purely photonic corrections, the
extreme accuracy of MUonE will presumably also demand for the
inclusion of the dominant effects beyond NNLO from fermionic
corrections. Among those, contributions due to electron pairs are the
most important. The evaluation of NLO photonic corrections to the
cross section $e\mu\to e\mu (e^+ e^-)$ combined with the corresponding
virtual electron-loop contributions will exhibit $\alpha^3 L_m^3$
terms. The determination of these terms could be achieved, for
example, by convoluting the NNLO cross section with a standard QED PS
simulation or by means of an appropriate generalization of the basic
ingredients of the PS algorithm. The resummation of pair production
contributions can be also shown to take place to all orders of the
perturbative expansion~\cite{Catani:1989et, Skrzypek:1992vk,
  Arbuzov:2010zzb}.

If necessary, many of the above estimates could be put on firmer
ground by computing the full set of virtual and real photon
corrections due to the radiation from a single leg at N$^3$LO
accuracy, as discussed in Section~\ref{sec:Bnnlo}.
 
To summarize, the accuracy of NNLO calculations combined with the
contributions due to multiple photon radiation will be limited by the
approximate inclusion of NLL contributions at ${\cal O} (\alpha^3)$. A
careful estimate of their impact on the observables measured by MUonE
will set the scale of the overall theoretical uncertainty.

\section{Monte Carlo} \label{montecarlo}

In the MUonE experiment, the extraction of the HVP contribution to the
effective electromagnetic coupling will be based on a template fitting
method. In this procedure, differential cross sections are calculated
according to a given theoretical input and compared to the data, as a
function of the parameters entering the $\Delta \alpha_{\rm had}
(q^2)$ modelling. Inevitably, this requires the implementation of the
theoretical predictions into a fully flexible MC code. The latter is
also needed for a high-precision calculation of the normalisation
cross section, as well as the evaluation of the detector efficiencies
and the assessment of a number of experimental systematics.  Thus, the
MC is the experimentally-oriented completion of any theory calculation
as it goes to the heart of the data analysis.

We describe here what is presently available in the sector of MC tools
for simulations of the $\mu$-$e$ scattering process and the most
important phenomenological results.  We are also interested in
providing a recipe on how to convert future theoretical achievements
or theory MC into useful tools for the experimentalists and
phenomenologists. A sketch of the ongoing efforts towards the
realisation of MC codes with increased accuracy or the simulation of
relevant contributions to $\mu$-$e$ scattering is also given.

Until now, feasibility studies and preliminary simulations by the MUonE
collaboration have been performed using a MC event generator that
includes NLO electroweak corrections.  The theoretical content of the
NLO MC is described in detail in \cite{Alacevich:2018vez} and will not
be repeated here.  Suffice it to say that the MC developed in
\cite{Alacevich:2018vez} is based on a calculation of the full set of
NLO electroweak corrections to $\mu$-$e$ scattering without any
approximation, including finite mass contributions.  More interesting
facts are the main computational features of the NLO MC. They can be
summarised as follows:
\begin{itemize}

\item 
    the generated events are fully exclusive, i.e. all the momenta of
    the event particles can be stored in such a way that {\it any}
    observable can be studied and {\it any} further effect can be
    applied (experimental cuts, detector simulation, etc);

\item 
    both {\it weighted} and {\it unweighted} (constant weight) events
    can be generated. The use of weighted events speeds up event
    generation and, generally, reduces the statistical error due to MC
    integration;

\item 
    the incoming muon energy (beam momentum) can be spread by a
    Gaussian distribution around its nominal va\-lue, to match realistic
    beam preparation;

\item 
    the HVP contribution can be switched on and off, all the rest of
    the input parameters remaining unchanged.  This gives the
    possibility of studying the contribution of $\Delta \alpha_{\rm
    had} (q^2)$ to any observable at NLO accuracy, including
    experimental effects\footnote{In the present version of the NLO
    MC, the HVP contribution is taken into account in terms of Fred
    Jegerlehner's routine {\tt
    hadr5n12}~\cite{Jegerlehner:2012}. Of
    course, any other available parametrisation can be easily
    interfaced and used.};

\item 
    the generated events can be stored into \textsc{Root} n-tuples for
    further analysis. The storage format includes all the relevant
    information for each run input and for each generated event. The
    flexible nature of the adopted format makes it suitable to
    facilitate the implementation of future theoretical developments,
    such as the inclusion of multiple photon emission;

\item 
    the code is equipped with a \textsc{Root} interface for reading,
    analysing and manipulating the generated samples.

\end{itemize}

In the following, we show a sample of particularly interesting
predictions obtained by means of the above NLO MC. Within the set of
numerical results described in \cite{Alacevich:2018vez}, we select
those that are particularly relevant in the light of the efforts in
the sector of NNLO corrections and resummation. To that purpose, we
are interested to address the following questions\footnote{We focus on
  photonic corrections, as the contribution of purely weak NLO
  corrections is well below the 10 ppm level, as shown in
  \cite{Alacevich:2018vez}.}:
\begin{itemize}
 
 \item 
    how the $\theta_e$-$\theta_\mu$ correlation of the elastic signal
    is affected by QED radiation at NLO and how the signal sensitivity
    can be recovered by applying suitable cuts;
 
 \item 
    how the full NLO QED correction is shared among the different
    gauge-invariant subsets described in Section~\ref{fixedorder};
 
 \item 
    how large finite electron-mass contributions are.

\end{itemize}
To answer the above questions, we provide numerical results for both
the $\mu^- e^- \to \mu^- e^-$ and $\mu^+ e^- \to \mu^+ e^-$ process,
since both options are relevant for the MUonE experiment and the mixed
QED corrections $\sim q^3Q^3$ differ in the two cases.

We use the following input parameters:
\begin{eqnarray}
  &\alpha(0) = 1/137.03599907430637&\nonumber\\
  &m= 0.510998928\text{ MeV} \qquad M = 105.6583715\text{ MeV}& 
\label{eq:inputqedparams}
\end{eqnarray}
where $\alpha(0)$ is the value used for the lepton-photon coupling.
For the energy of the incoming muons, we assume $E_1=150$~GeV, which
is the energy of the M2 beam line of the CERN SPS. Note that, under
the fixed-target configuration of the MUonE experiment, the CMS energy
corresponding to this muon energy is given by $\sqrt{s}\simeq
0.405541$~GeV and that the Lorentz $\gamma$ factor boosting from CMS
to LAB is $\gamma\simeq 370$.  Due to \Eqn{eq:te}, a lower limit on
$E_4$ implies an upper limit on $t_{e}$. In this kinematical
condition, the collinear logarithms $L_e = \ln (|t_{\rm max}|/m^2)$
and $L_\mu = \ln (|t_{\rm max}|/M^2)$ amount to $L_e \simeq 13.4$ and
$L_\mu \simeq 2.7$, respectively.

To study the dependence of the radiative corrections on the applied
cuts, we consider {two different event selections defined by the
following criteria}:
\begin{enumerate}

\item 
    $\theta_e,\theta_\mu<100\mathrm{\ mrad}$ and $E_4 > 0.2$~GeV
    {(i.e. $t_{e}\lesssim -2.04\cdot 10^{-4}\text{ GeV}^2$)}. The
    angular cuts model the typical acceptance conditions of MUonE and
    the electron energy threshold is imposed to guarantee the presence
    of two charged tracks in the detector;

\item 
    the same criteria as in Setup 1, with an additional acoplanarity
    cut, applied to partially remove radiative events and thus
    enhancing the fraction of elastic events. We require
    acoplanarity~$\big|\pi - |\phi_e-\phi_\mu|\big|$ lower than
    $3.5\mathrm{\ mrad}$, for the sake of illustration.

\end{enumerate}

\begin{figure}
\begin{center}
\includegraphics[width=\columnwidth]{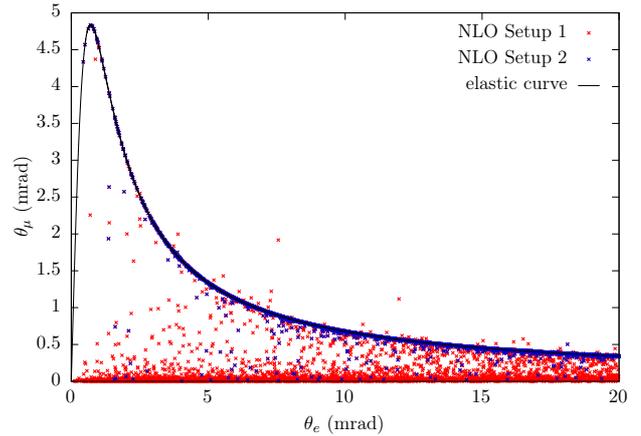}
\end{center}
\caption{The correlation between the electron
scattering angle $\theta_e$ and muon scattering angle $\theta_\mu$ for
the $\mu^+ e^- \to \mu^+ e^-$ process at LO (elastic curve) and NLO
QED, for the selection criteria 1 and 2 defined in the text.}
\label{Fig:Figure1-MC} 
\end{figure}

The answer to the first question about the impact of NLO QED radiation
on the $\theta_e$-$\theta_\mu$ elastic correlation is given by
Figure~\ref{Fig:Figure1-MC}. In that figure, we compare the
correlation in the laboratory frame between the scattering angles of
the outgoing electron and muon at LO and NLO, for the Setup 1 and
Setup 2 defined above. It can be noticed that, in the absence of an
acoplanarity cut (Setup 1), the correlation present at LO (elastic
curve) is largely modified by the presence of events at relatively
small muon angles, which originate from the bremsstrahlung process
$\mu^+ e^- \to \mu^+ e^-\gamma$.  However, the tight acoplanarity cut
(Setup 2) turns out to be effective in getting rid of most of these
radiative events, thus isolating the elastic {correlation curve}. As
shown in \cite{Alacevich:2018vez}, in the presence of acceptance cuts
only (Setup 1), the corrections to the electron scattering angle turn
out to be quite sizeable at small angles, due to the emission of a
hard photon in the radiative process $\mu e \to \mu e \gamma$.
However, this effect gets largely reduced when an elasticity cut is
applied (Setup 2), yielding a correction in the 10-40\% range for all
the relevant distributions. In the presence of an elasticity cut that
vetoes hard photon emission, the contribution of soft photons becomes
enhanced and gives rise to large IR logarithms, as remarked in
Section~\ref{resummation}.  Not to invalidate the perturbative
expansion, those logarithms need to be resummed together with the
contributions due to collinear emission. This can be achieved by means
of exclusive MC techniques, such as YFS exponentiation or QED Parton
Shower, or analytic resummation. As emphasised in
Section~\ref{resummation}, the latter method requires the
identification of a kinematical quantity able to select elastic
scattering. This poses the question how the elasticity band isolated
by the cuts of Setup 2 can be approximated by a reasonably simple
`observable' suitable for resummation.


\begin{figure}[ht]
\vspace{2mm}
\begin{center}
\subfloat[$d\sigma/dt_e$]{\includegraphics[width=\columnwidth]{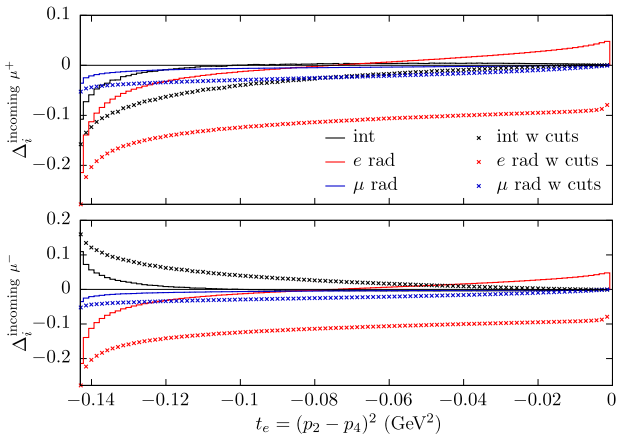}}
\vspace{1mm}

\subfloat[$d\sigma/dt_\mu$]{\includegraphics[width=\columnwidth]{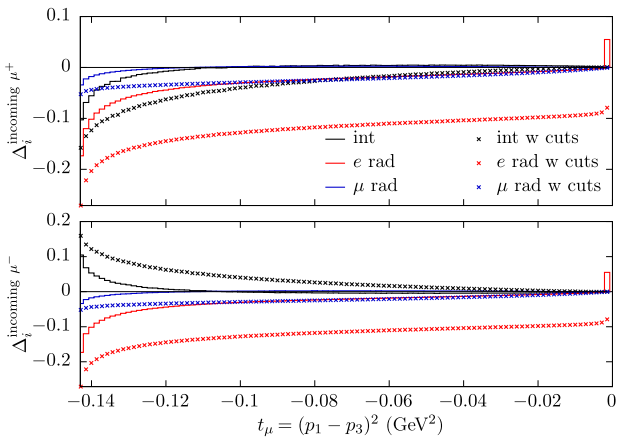}}
\end{center}
\caption{The contribution of the QED gauge-invariant subsets to the
cross section of the process $\mu^+ e^- \to \mu^+ e^-$ (upper panel)
and of the process $\mu^- e^- \to \mu^- e^-$ (lower panel), as a
function of the squared momentum transfer $t_e$ and $t_\mu$. The
results refer to Setup 1 (solid lines) and Setup 2 (dotted lines)
described in the text.}
\label{Fig:Figure2-MC} 
\end{figure}

To understand how the gauge-invariant subsets contribute to the
overall NLO QED correction, we show in Figure~\ref{Fig:Figure2-MC} the
impact of the different classes described Section~\ref{fixedorder} on
the $d\sigma / d t_{e}$ (top plot) and $d\sigma / d t_{\mu}$ (bottom
plot) distributions.  The upper (lower) panels refer to the $\mu^+ e^-
\to \mu^+ e^-$($\mu^- e^- \to \mu^- e^-$) process. The squared
momentum transfers $t_{e}$ and $t_{\mu}$ are defined in \Eqn{eq:te}
and \Eqn{eq:tm}, respectively.  The main message that can be drawn
from Figure~\ref{Fig:Figure2-MC} is that the NLO QED correction over
the full range is, in general, the result of a subtle interplay
between the various sources of radiation.  A further general remark is
that the mixed corrections due to electron-muon interference are of
opposite sign for the two processes and particularly relevant for
large $|t_{\mu,e}|$ values. The latter behaviour has to be ascribed to
the presence in the up-down interference of logarithmic (and squared
logarithmic) angular contributions of the type $\ln(u/t)$, which
become potentially enhanced when either $t$ or $u$ are small. More in
detail, one can see from Figure~\ref{Fig:Figure2-MC} that, in the
presence of acceptance cuts only, the NLO correction is dominated by
the contribution of electron radiation, the other effects being almost
flat and much smaller over the full range. However, if an acoplanarity
cut is applied, the contributions due to muon radiation and up-down
interference corrections become visible for large $|t|$ values, where
they amount to some percent. Interestingly, the above contributions
have the same sign in the $\mu^+ e^- \to \mu^+ e^-$ process (upper
panel of the left plot) and sum up to contribute to the overall QED
correction, whereas they tend to cancel in the $\mu^- e^- \to \mu^-
e^-$ process (lower panel of the left plot). Therefore, also in view
of ongoing calculations at NNLO, these results indicate that all the
gauge-invariant subsets have to be taken into account.
{\hbadness=10000

\begin{figure}
\vspace{2mm}
\begin{center}
\subfloat[$d\sigma/d\theta_e$]{\includegraphics[width=\columnwidth]{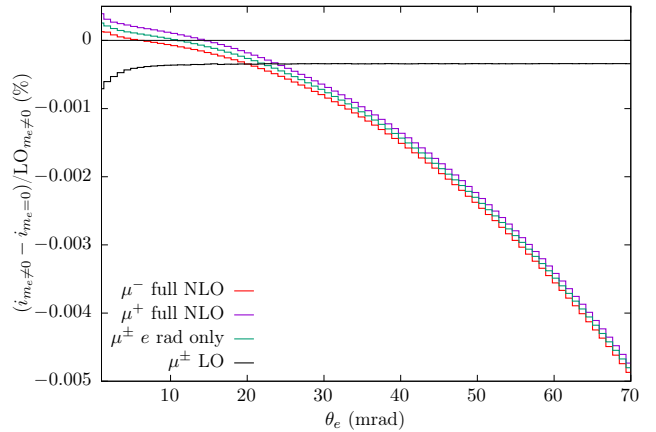}}

\subfloat[$d\sigma/dt_e$]{\includegraphics[width=\columnwidth]{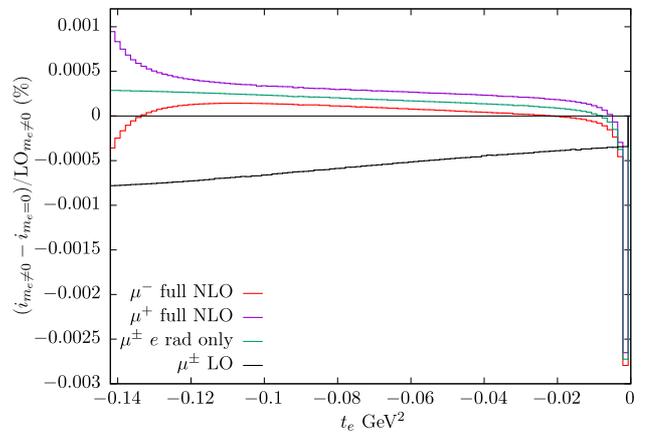}}
\end{center}
\caption{The relative contribution of
finite electron-mass corrections to the cross section of the processes
$\mu^\pm e^- \to \mu^\pm e^-$, as a function of the electron
scattering angle and the momentum transfer.  The predictions
refer to Setup 1 defined in the text.}
\vspace{\baselineskip}
\label{Fig:Figure3-MC} 
\end{figure}

}The size of finite electron-mass contributions at NLO is illustrated
in Figure~\ref{Fig:Figure3-MC}, where the results are shown in
per-cent of the fully massive LO differential cross sections.  The
predictions refer to both incoming $\mu^+$ and $\mu^-$ and are shown
for the electron scattering angle and the squared momentum transfer
$t_{e}$, for the sake of illustration. Similar results hold for other
distributions. A sensible assessment of mass contributions beyond
logarithmic accuracy is in general a delicate issue for any
fixed-order calculation and particularly tricky for $\mu$-$e$
scattering under MUonE conditions, where the limit of massless
electron implies that its rest frame (i.e. the lab frame) can not be
defined.  To bypass this difficulty, we follow the procedure detailed
in \cite{Alacevich:2018vez}, that allows to get an estimate of finite
electron-mass contributions for the sum of one-loop virtual and real
soft-photon corrections. The one-loop virtual amplitude is split into
a contribution that is proportional to logarithms of the (artificial
photon mass) IR parameter and the remainder, ${\cal M}_{n}^{(1)} =
{\cal M}_{n, {\rm IR}}^{(1)} + {\cal M}_{n, {\rm non \, IR}}^{(1)}
$. A similar split is done for the amplitude related to soft real
emission.  According to the notation of Section~\ref{fixedorder}, the
$m \to 0$ limit of the NLO correction is then evaluated according to
the chain formula
\begin{align}
  \bigg[& {\cal M}_{n, {\rm IR}}^{(1)} (m)
    + {\cal M}_{n, {\rm non \, IR}}^{(1)} (zM)|_{z=0} \nonumber \\
    &+ {\cal M}_{n+1, {\rm IR}}^{(0)/{\rm soft}} (m) 
    + {\cal M}_{n+1, {\rm non \, IR}}^{(0)/{\rm soft}} (zM)|_{z=0}
    \bigg] \times d\Phi_n(m) \nonumber\\
 &+ {\cal M}_{n+1}^{(0)/{\rm hard}} (m) \times d\Phi_{n+1}(m)
\end{align}
that provides an IR-safe estimate of electron mass contributions,
while keeping exact kinematics and phase space.

As can be seen from Figure~\ref{Fig:Figure3-MC}, the contribution of
$m$-dependent terms to the LO cross section is almost flat and below
the 10 ppm level. The electron-mass corrections at NLO contribute to
$d\sigma / d\theta_e$ and $d\sigma / d t_{e}$ in the range from a few
to some $10^{-5}$. We notice that the largest part of the finite $m$
corrections is due to radiation from the electron line only, the full
correction lying around it. The extra corrections w.r.t. electron line
only are dominated by up-down interference and box diagrams. These
results suggest that electron mass contributions beyond logarithmic
accuracy can be neglected in a NNLO computation or, eventually,
included at the level of electron line corrections only.  Actually, a
rescaling of the first-order contribution shown in
Figure~\ref{Fig:Figure3-MC}, which is at most of the order of
$10^{-5}$, by a factor $(\alpha/\pi) \ln(-t/m^2)$ provides an estimate
of the electron mass effects at NNLO and yields a correction much
smaller than 10\ppm.

The presently available MC at NLO accuracy represents just the first
step towards the realisation of a high-precision theoretical tool
necessary for the data analysis of $\mu$-$e$ scattering by MUonE. The
ultimate goal is the realisation of a MC code including NNLO
corrections and resummation of QED contributions due to multiple
photon radiation. However, over a relatively short term, a number of
intermediate results could be obtained about some important
contributions beyond NLO.

A first example is given by the matching of NLO corrections to a QED
parton shower, following the formulation already applied to Bhabha
scattering and $e^+ e^-$ annihilation processes in QED
\cite{Balossini:2006wc,Balossini:2008xr}, Drell-Yan processes
\cite{CarloniCalame:2006zq,CarloniCalame:2007cd} and Higgs decay into
four leptons \cite{Boselli:2015aha}. This would allow to estimate the
most relevant QED corrections beyond NLO under realistic event
selection criteria.

A further prospect under consideration is the calculation of lepton
pair corrections to $\mu$-$e$ scattering. These corrections appear at
NNLO and are a combination of two-loop virtual lepton-loop corrections
with the same-order contribution of real pair emission, i.e. $\mu e
\to \mu e + (\ell^+ \ell^-)$, with $\ell = e, \mu$. One and two-loop
diagrams with vacuum polarisation insertions in the photon propagator
were considered some time ago for the case of the Bhabha scattering in
the massless limit with the 0.1\%
accuracy~\cite{Arbuzov:1995qd,Arbuzov:1995vj, Arbuzov:1995cn,
  Arbuzov:1995vi,Montagna:1998vb, CarloniCalame:2011zq}.

Such a calculation can be extended to the treatment of hadronic pair
corrections, by combining the already available virtual hadronic
contributions \cite{Fael:2019nsf} with the process of pion pair
production $\mu e \to \mu e + (\pi^0 \pi^0, \pi^+ \pi^-)$.  A further
step that can be taken is the evaluation of the background process
$\mu e \to \mu e + (\pi^0 \to \gamma\gamma)$, that could benefit, as
for two-pion production, from the experience in the development of MC
generators for the simulation of hadronic final states at flavor
factories \cite{Actis:2010gg}. Finally, one should not forget that
electrons are bound inside the target and the impact of bound-state
effects should be evaluated. This will require considering the
possibility of scattering of the incident muons off core valence
electrons, for which off-shell effects due to the finite binding
energy and momentum distribution must be considered.

All the above developments are under consideration and preliminary
results are also available for most of them.

\section{Summary} \label{summary}

With this report we want to document that within the theory community
there is sufficient interest, manpower, and expertise to provide the
necessary theory support for the MUonE experiment.

The minimal goal that will be achieved in a first step is a fully
differential parton-level Monte Carlo program containing the following
contributions: (i) the fully massive NLO QED (and electroweak)
contributions; (ii) the fully massive (dominant) electronic
contributions at NNLO; (iii) the fully massive NNLO hadronic contributions;
(iv) the remaining contributions at NNLO in a massified approach,
i.e. neglecting finite electron mass terms. In addition, this
fixed-order calculation will be matched to a parton shower taking into
account multiple photon emission at leading logarithmic accuracy.

Given the current status described in this report, there are no
further conceptual challenges that need to be overcome to achieve
this. Needless to say that nevertheless there will still be numerous
difficult issues to be sorted. Hence, ideally there will be at least
two different implementations of such a code to facilitate
debugging. To that end, the theory groups of Pavia and PSI are both
committed to each produce an implementation.

{\hbadness=10000 In} a second step, a detailed, realistic
phenomenological analysis is required to investigate if this theory
description is sufficient. A careful error estimate of the missing
terms will be crucial. This analysis will be done in close
collaboration with the experimental collaboration.

It is quite likely that a third step will be required, i.e. further
improvements to the theory. Most probably, the next-to-leading
logarithms will have to be addressed. A careful study of the
logarithms related to the factorisation anomaly is also important. In
connection to fixed-order calculations, it is not at all unrealistic
to expect a fully differential N$^3$LO computation of the dominant
electronic contributions in time for the MUonE experiment. Also, a
complete fully-massive NNLO calculation, possibly using numerical
techniques, is a serious target for theory in the longer term.

In fact, all the theory questions that are to be addressed in
connection with $\mu$-$e$ scattering are also of interest to a much
wider community. The developments that are made in this -- from a
theory point of view -- simple framework will undoubtedly lead to
progress in related fields. Thus, apart from providing an alternative
determination of the HVP, MUonE can also act as an icebreaker to free
a path for further theory progress.

\bigskip
\overfullrule=1mm
\section*{Acknowledgements}
We would like to thank the Physik-Insitut of UZH for its hospitality
in hosting the 2$^\text{nd}$ WorkStop/ThinkStart.  A special thanks to
Monika R\"{o}llin for her invaluable help in organising this
event.

We would also like to thank the University of Padova and INFN Padova
and the MITP for their hospitality and support of the 2017 Theory
Kickoff Workshop and 2018 Topical Workshop, respectively.

P.M. and W.J.T. wish to thank Giulio Dondi for testing the software
AIDA on the real-virtual NNLO calculations.
P.B. acknowledges support by the European Union's Horizon 2020 research
and innovation programme under the Marie Sk\l{}odowska-Curie grant
agreement No 701647.
The work of M.C. is supported by the Investissements d'avenir, Labex
ENIGMASS.
T.E. and Y.U. are supported by the Swiss National Science Foundation
under contract 200021\_178967.  Y.U. is partially supported by a
Forschungskredit of the University of Zurich under contract number
FK-19-087.
M.F. is supported by the Deutsche Forschungsgemeinschaft (DFG,
German Research Foundation) under grant 396021762 - TRR 257 ``Particle
Physics Phenomenology after the Higgs Discovery''.
The work of P.M. is supported by the grant Supporting TAlent in
ReSearch at Padova University (UniPD STARS Grant 2017
``Diagrammalgebra'').
W.J.T. is partially supported by Grants No. FPA2017-84445-P and
No. SEV-2014-0398 (AEI/ERDF, EU), the COST Action CA16201
PARTICLEFACE, and the ``Juan de la Cierva Formaci\'on'' program
(FJCI-2017-32128).
G.O. was supported in part by PSC-CUNY Award 61155-00 49.
M.P. acknowledges partial support by FP10 ITN Elusives
(H2020-MSCA-ITN-2015-674896) as well as Invisibles-Plus
(H2020-MSCA-RISE-2015-690575).
G.V. acknowledges the  EU STRONG 2020 project (Grant Agreement
824093).

{\hbadness=10000
\raggedbottom
\bibliographystyle{JHEP}
\bibliography{muone_ref}

\providecommand{\href}[2]{#2}\begingroup\raggedright\begin{thebibliography}{100}

\bibitem{Abbiendi:2016xup}
G.~Abbiendi et~al., \emph{{Measuring the leading hadronic contribution to the
  muon $g-2$ via $\mu e$ scattering}},
  \href{https://doi.org/10.1140/epjc/s10052-017-4633-z}{\emph{Eur. Phys. J.}
  {\bfseries C77} (2017) 139}
  [\href{https://arxiv.org/abs/1609.08987}{{\ttfamily 1609.08987}}].

\bibitem{LoI}
{\scshape MUonE} collaboration, \emph{{The MUonE Project}}, {\emph{Letter of
  Intent} {\bfseries CERN-SPSC-2019-026 / SPSC-I-252} (2019) }.

\bibitem{Calame:2015fva}
C.~M. Carloni~Calame, M.~Passera, L.~Trentadue and G.~Venanzoni, \emph{{A new
  approach to evaluate the leading hadronic corrections to the muon $g-2$}},
  \href{https://doi.org/10.1016/j.physletb.2015.05.020}{\emph{Phys. Lett.}
  {\bfseries B746} (2015) 325}
  [\href{https://arxiv.org/abs/1504.02228}{{\ttfamily 1504.02228}}].

\bibitem{Arbuzov:2004wp}
A.~B. Arbuzov, D.~Haidt, C.~Matteuzzi, M.~Paganoni and L.~Trentadue, \emph{{The
  Running of the electromagnetic coupling $\alpha$ in small angle Bhabha
  scattering}}, \href{https://doi.org/10.1140/epjc/s2004-01649-0}{\emph{Eur.
  Phys. J.} {\bfseries C34} (2004) 267}
  [\href{https://arxiv.org/abs/hep-ph/0402211}{{\ttfamily hep-ph/0402211}}].

\bibitem{Abbiendi:2005rx}
{\scshape OPAL} collaboration, G.~Abbiendi et~al., \emph{{Measurement of the
  running of the QED coupling in small-angle Bhabha scattering at LEP}},
  \href{https://doi.org/10.1140/epjc/s2005-02389-3}{\emph{Eur. Phys. J. C}
  {\bfseries 45} (2006) 1}
  [\href{https://arxiv.org/abs/hep-ex/0505072}{{\ttfamily hep-ex/0505072}}].

\bibitem{Abbiendi:2019qtw}
G.~Abbiendi et~al., \emph{{Results on Multiple Coulomb Scattering from 12 and
  20 GeV electrons on Carbon targets}},
  \href{https://doi.org/10.1088/1748-0221/15/01/P01017}{\emph{JINST} {\bfseries
  15} (2020) P01017} [\href{https://arxiv.org/abs/1905.11677}{{\ttfamily
  1905.11677}}].

\bibitem{Masiero:2020vxk}
A.~Masiero, P.~Paradisi and M.~Passera, \emph{{New physics at the MUonE
  experiment at CERN}},  \href{https://arxiv.org/abs/2002.05418}{{\ttfamily
  2002.05418}}.

\bibitem{Dev:2020drf}
P.~S.~B. Dev, W.~Rodejohann, X.-J. Xu and Y.~Zhang, \emph{{MUonE sensitivity to
  new physics explanations of the muon anomalous magnetic moment}},
  \href{https://arxiv.org/abs/2002.04822}{{\ttfamily 2002.04822}}.

\bibitem{Jegerlehner:2001ca}
F.~Jegerlehner, \emph{{The Effective fine structure constant at TESLA
  energies}},  \href{https://arxiv.org/abs/hep-ph/0105283}{{\ttfamily
  hep-ph/0105283}}.

\bibitem{Jegerlehner:2006ju}
F.~Jegerlehner, \emph{{Precision measurements of $\sigma_\text{hadronic}$ for
  $\alpha_\text{eff}(E)$ at ILC energies and $(g-2)_\mu$}},
  \href{https://doi.org/10.1016/j.nuclphysbps.2006.09.060}{\emph{Nucl. Phys.
  Proc. Suppl.} {\bfseries 162} (2006) 22}
  [\href{https://arxiv.org/abs/hep-ph/0608329}{{\ttfamily hep-ph/0608329}}].

\bibitem{Jegerlehner:2011mw}
F.~Jegerlehner, \emph{{Electroweak effective couplings for future precision
  experiments}}, \href{https://doi.org/10.1393/ncc/i2011-11011-0}{\emph{Nuovo
  Cim.} {\bfseries C034S1} (2011) 31}
  [\href{https://arxiv.org/abs/1107.4683}{{\ttfamily 1107.4683}}].

\bibitem{Harlander:2002ur}
R.~V. Harlander and M.~Steinhauser, \emph{{{\tt rhad}: A Program for the
  evaluation of the hadronic R ratio in the perturbative regime of QCD}},
  \href{https://doi.org/10.1016/S0010-4655(03)00204-2}{\emph{Comput. Phys.
  Commun.} {\bfseries 153} (2003) 244}
  [\href{https://arxiv.org/abs/hep-ph/0212294}{{\ttfamily hep-ph/0212294}}].

\bibitem{Bardin:1997nc}
D.~{\relax Yu}. Bardin and L.~Kalinovskaya, \emph{{QED corrections for
  polarized elastic $\mu e$ scattering}},
  \href{https://arxiv.org/abs/hep-ph/9712310}{{\ttfamily hep-ph/9712310}}.

\bibitem{Kaiser:2010zz}
N.~Kaiser, \emph{{Radiative corrections to lepton-lepton scattering
  revisited}}, \href{https://doi.org/10.1088/0954-3899/37/11/115005}{\emph{J.
  Phys.} {\bfseries G37} (2010) 115005}.

\bibitem{Alacevich:2018vez}
M.~Alacevich, C.~M. Carloni~Calame, M.~Chiesa, G.~Montagna, O.~Nicrosini and
  F.~Piccinini, \emph{{Muon-electron scattering at NLO}},
  \href{https://doi.org/10.1007/JHEP02(2019)155}{\emph{JHEP} {\bfseries 02}
  (2019) 155} [\href{https://arxiv.org/abs/1811.06743}{{\ttfamily
  1811.06743}}].

\bibitem{NLOmfmp}
M.~Fael and M.~Passera, \emph{private communication}, {\emph{unpublished}
  (2018) }.

\bibitem{NLOus}
T.~Engel, A.~Signer and Y.~Ulrich, \emph{private communication},
  {\emph{unpublished} (2019) }.

\bibitem{Bonciani:2003ai}
R.~Bonciani, P.~Mastrolia and E.~Remiddi, \emph{{QED vertex form-factors at two
  loops}}, \href{https://doi.org/10.1016/j.nuclphysb.2003.10.031}{\emph{Nucl.
  Phys.} {\bfseries B676} (2004) 399}
  [\href{https://arxiv.org/abs/hep-ph/0307295}{{\ttfamily hep-ph/0307295}}].

\bibitem{Bernreuther:2004ih}
W.~Bernreuther, R.~Bonciani, T.~Gehrmann, R.~Heinesch, T.~Leineweber et~al.,
  \emph{{Two-loop QCD corrections to the heavy quark form-factors: The Vector
  contributions}},
  \href{https://doi.org/10.1016/j.nuclphysb.2004.10.059}{\emph{Nucl.Phys.}
  {\bfseries B706} (2005) 245}
  [\href{https://arxiv.org/abs/hep-ph/0406046}{{\ttfamily hep-ph/0406046}}].

\bibitem{Engel:2019nfw}
T.~Engel, A.~Signer and Y.~Ulrich, \emph{{A subtraction scheme for massive
  QED}}, \href{https://doi.org/10.1007/JHEP01(2020)085}{\emph{JHEP} {\bfseries
  02} (2020) 085} [\href{https://arxiv.org/abs/1909.10244}{{\ttfamily
  1909.10244}}].

\bibitem{Frixione:1995ms}
S.~{Frixione}, Z.~{Kunszt} and A.~{Signer}, \emph{Three-jet cross sections to
  next-to-leading order}, {\emph{Nuclear Physics B} {\bfseries 467} (1996) 399}
  [\href{https://arxiv.org/abs/hep-ph/9512328v1}{{\ttfamily
  hep-ph/9512328v1}}].

\bibitem{Frederix:2009yq}
R.~{Frederix}, S.~{Frixione}, F.~{Maltoni} and T.~{Stelzer}, \emph{{Automation
  of next-to-leading order computations in QCD: the FKS subtraction}},
  {\emph{Journal of High Energy Physics} {\bfseries 2009} (2009) }
  [\href{https://arxiv.org/abs/0908.4272v2}{{\ttfamily 0908.4272v2}}].

\bibitem{Engel:2019}
T.~Engel, P.~Banerjee, A.~Signer and Y.~Ulrich, \emph{{NNLO corrections in
  massive QED}},
  \href{https://indico.psi.ch/event/6857/contributions/18942/}{https://indico.psi.ch/event/6857/contributions/18942/}.

\bibitem{Ulrich:2019}
Y.~Ulrich, \emph{{High-precision QED prediction for low-energy lepton
  experiments}},
  \href{https://indico.psi.ch/event/6857/contributions/19673/}{https://indico.psi.ch/event/6857/contributions/19673/}.

\bibitem{Bucoveanu:2018soy}
R.~D. Bucoveanu and H.~Spiesberger, \emph{{Second-Order Leptonic Radiative
  Corrections for Lepton-Proton Scattering}},
  \href{https://doi.org/10.1140/epja/i2019-12727-1}{\emph{Eur. Phys. J.}
  {\bfseries A55} (2019) 57}
  [\href{https://arxiv.org/abs/1811.04970}{{\ttfamily 1811.04970}}].

\bibitem{Dondi:2019th}
G.~Dondi, \emph{{Unitarity-based Methods for Muon-Electron Scattering in
  Quantum Electrodynamics}},  Master's thesis, Padua U., 2019.

\bibitem{Hahn:2000kx}
T.~Hahn, \emph{{Generating Feynman diagrams and amplitudes with FeynArts 3}},
  \href{https://doi.org/10.1016/S0010-4655(01)00290-9}{\emph{Comput.Phys.Commun.}
  {\bfseries 140} (2001) 418}
  [\href{https://arxiv.org/abs/hep-ph/0012260}{{\ttfamily hep-ph/0012260}}].

\bibitem{Shtabovenko:2016sxi}
V.~Shtabovenko, R.~Mertig and F.~Orellana, \emph{{New Developments in FeynCalc
  9.0}}, \href{https://doi.org/10.1016/j.cpc.2016.06.008}{\emph{Comput. Phys.
  Commun.} {\bfseries 207} (2016) 432}
  [\href{https://arxiv.org/abs/1601.01167}{{\ttfamily 1601.01167}}].

\bibitem{Hodges:2009hk}
A.~Hodges, \emph{{Eliminating spurious poles from gauge-theoretic amplitudes}},
  \href{https://doi.org/10.1007/JHEP05(2013)135}{\emph{JHEP} {\bfseries 05}
  (2013) 135} [\href{https://arxiv.org/abs/0905.1473}{{\ttfamily 0905.1473}}].

\bibitem{Badger:2016uuq}
S.~Badger, \emph{{Automating QCD amplitudes with on-shell methods}},
  \href{https://doi.org/10.1088/1742-6596/762/1/012057}{\emph{J. Phys. Conf.
  Ser.} {\bfseries 762} (2016) 012057}
  [\href{https://arxiv.org/abs/1605.02172}{{\ttfamily 1605.02172}}].

\bibitem{Driencourt-Mangin:2019yhu}
F.~Driencourt-Mangin, G.~Rodrigo, G.~F.~R. Sborlini and W.~J. Torres~Bobadilla,
  \emph{{On the interplay between the loop-tree duality and helicity
  amplitudes}},  \href{https://arxiv.org/abs/1911.11125}{{\ttfamily
  1911.11125}}.

\bibitem{Tkachov:1981wb}
F.~V. Tkachov, \emph{{A Theorem on Analytical Calculability of Four Loop
  Renormalization Group Functions}},
  \href{https://doi.org/10.1016/0370-2693(81)90288-4}{\emph{Phys. Lett.}
  {\bfseries 100B} (1981) 65}.

\bibitem{Chetyrkin:1981qh}
K.~G. Chetyrkin and F.~V. Tkachov, \emph{{Integration by Parts: The Algorithm
  to Calculate $\beta$-functions in 4 Loops}},
  \href{https://doi.org/10.1016/0550-3213(81)90199-1}{\emph{Nucl. Phys.}
  {\bfseries B192} (1981) 159}.

\bibitem{Laporta:2001dd}
S.~Laporta, \emph{{High precision calculation of multiloop Feynman integrals by
  difference equations}}, \href{https://doi.org/10.1016/S0217-751X(00)00215-7,
  10.1142/S0217751X00002157}{\emph{Int. J. Mod. Phys.} {\bfseries A15} (2000)
  5087} [\href{https://arxiv.org/abs/hep-ph/0102033}{{\ttfamily
  hep-ph/0102033}}].

\bibitem{Maierhoefer:2017hyi}
P.~Maierh{\"o}fer, J.~Usovitsch and P.~Uwer, \emph{{Kira---A Feynman integral
  reduction program}},
  \href{https://doi.org/10.1016/j.cpc.2018.04.012}{\emph{Comput. Phys. Commun.}
  {\bfseries 230} (2018) 99}
  [\href{https://arxiv.org/abs/1705.05610}{{\ttfamily 1705.05610}}].

\bibitem{Mastrolia:2016dhn}
P.~Mastrolia, T.~Peraro and A.~Primo, \emph{{Adaptive Integrand Decomposition
  in parallel and orthogonal space}},
  \href{https://doi.org/10.1007/JHEP08(2016)164}{\emph{JHEP} {\bfseries 08}
  (2016) 164} [\href{https://arxiv.org/abs/1605.03157}{{\ttfamily
  1605.03157}}].

\bibitem{Mastrolia:2016czu}
P.~Mastrolia, T.~Peraro, A.~Primo and W.~J. Torres~Bobadilla, \emph{{Adaptive
  Integrand Decomposition}},
  \href{https://doi.org/10.22323/1.260.0007}{\emph{PoS} {\bfseries LL2016}
  (2016) 007} [\href{https://arxiv.org/abs/1607.05156}{{\ttfamily
  1607.05156}}].

\bibitem{AIDA}
P.~Mastrolia, T.~Peraro, A.~Primo, W.~J. Torres~Bobadilla, L.~Mattiazzi,
  J.~Ronca et~al., \emph{{AIDA: adaptive integrand decomposition algorithm}},
  {\emph{{\rm Private version}} }.

\bibitem{vonManteuffel:2012np}
A.~von Manteuffel and C.~Studerus, \emph{{Reduze 2 - Distributed Feynman
  Integral Reduction}},  \href{https://arxiv.org/abs/1201.4330}{{\ttfamily
  1201.4330}}.

\bibitem{vonManteuffel:2014qoa}
A.~von Manteuffel, E.~Panzer and R.~M. Schabinger, \emph{{A quasi-finite basis
  for multi-loop Feynman integrals}},
  \href{https://doi.org/10.1007/JHEP02(2015)120}{\emph{JHEP} {\bfseries 02}
  (2015) 120} [\href{https://arxiv.org/abs/1411.7392}{{\ttfamily 1411.7392}}].

\bibitem{Mastrolia:2017pfy}
P.~Mastrolia, M.~Passera, A.~Primo and U.~Schubert, \emph{{Master integrals for
  the NNLO virtual corrections to $\mu e$ scattering in QED: the planar
  graphs}}, \href{https://doi.org/10.1007/JHEP11(2017)198}{\emph{JHEP}
  {\bfseries 11} (2017) 198}
  [\href{https://arxiv.org/abs/1709.07435}{{\ttfamily 1709.07435}}].

\bibitem{DiVita:2018nnh}
S.~Di~Vita, S.~Laporta, P.~Mastrolia, A.~Primo and U.~Schubert, \emph{{Master
  integrals for the NNLO virtual corrections to $\mu e$ scattering in QED: the
  non-planar graphs}},
  \href{https://doi.org/10.1007/JHEP09(2018)016}{\emph{JHEP} {\bfseries 09}
  (2018) 016} [\href{https://arxiv.org/abs/1806.08241}{{\ttfamily
  1806.08241}}].

\bibitem{Barucchi:1973zm}
G.~Barucchi and G.~Ponzano, \emph{{Differential equations for one-loop
  generalized feynman integrals}},
  \href{https://doi.org/10.1063/1.1666327}{\emph{J. Math. Phys.} {\bfseries 14}
  (1973) 396}.

\bibitem{Kotikov:1990kg}
A.~V. Kotikov, \emph{{Differential equations method: New technique for massive
  Feynman diagrams calculation}},
  \href{https://doi.org/10.1016/0370-2693(91)90413-K}{\emph{Phys. Lett.}
  {\bfseries B254} (1991) 158}.

\bibitem{Remiddi:1997ny}
E.~Remiddi, \emph{{Differential equations for Feynman graph amplitudes}},
  {\emph{Nuovo Cim.} {\bfseries A110} (1997) 1435}
  [\href{https://arxiv.org/abs/hep-th/9711188}{{\ttfamily hep-th/9711188}}].

\bibitem{Gehrmann:1999as}
T.~Gehrmann and E.~Remiddi, \emph{{Differential equations for two loop four
  point functions}},
  \href{https://doi.org/10.1016/S0550-3213(00)00223-6}{\emph{Nucl. Phys.}
  {\bfseries B580} (2000) 485}
  [\href{https://arxiv.org/abs/hep-ph/9912329}{{\ttfamily hep-ph/9912329}}].

\bibitem{Henn:2013pwa}
J.~M. Henn, \emph{{Multiloop integrals in dimensional regularization made
  simple}}, \href{https://doi.org/10.1103/PhysRevLett.110.251601}{\emph{Phys.
  Rev. Lett.} {\bfseries 110} (2013) 251601}
  [\href{https://arxiv.org/abs/1304.1806}{{\ttfamily 1304.1806}}].

\bibitem{Argeri:2014qva}
M.~Argeri, S.~Di~Vita, P.~Mastrolia, E.~Mirabella, J.~Schlenk, U.~Schubert
  et~al., \emph{{Magnus and Dyson Series for Master Integrals}},
  \href{https://doi.org/10.1007/JHEP03(2014)082}{\emph{JHEP} {\bfseries 03}
  (2014) 082} [\href{https://arxiv.org/abs/1401.2979}{{\ttfamily 1401.2979}}].

\bibitem{DiVita:2014pza}
S.~Di~Vita, P.~Mastrolia, U.~Schubert and V.~Yundin, \emph{{Three-loop master
  integrals for ladder-box diagrams with one massive leg}},
  \href{https://doi.org/10.1007/JHEP09(2014)148}{\emph{JHEP} {\bfseries 09}
  (2014) 148} [\href{https://arxiv.org/abs/1408.3107}{{\ttfamily 1408.3107}}].

\bibitem{DiVita:2019lpl}
S.~Di~Vita, T.~Gehrmann, S.~Laporta, P.~Mastrolia, A.~Primo and U.~Schubert,
  \emph{{Master integrals for the NNLO virtual corrections to $
  q\overline{q}\to t\overline{t} $ scattering in QCD: the non-planar graphs}},
  \href{https://doi.org/10.1007/JHEP06(2019)117}{\emph{JHEP} {\bfseries 06}
  (2019) 117} [\href{https://arxiv.org/abs/1904.10964}{{\ttfamily
  1904.10964}}].

\bibitem{Bauer:2000cp}
C.~W. Bauer, A.~Frink and R.~Kreckel, \emph{{Introduction to the GiNaC
  framework for symbolic computation within the C++ programming language}},
  {\emph{J. Symb. Comput.} {\bfseries 33} (2000) 1}
  [\href{https://arxiv.org/abs/cs/0004015}{{\ttfamily cs/0004015}}].

\bibitem{Borowka:2015mxa}
S.~Borowka, G.~Heinrich, S.~P. Jones, M.~Kerner, J.~Schlenk and T.~Zirke,
  \emph{{SecDec-3.0: numerical evaluation of multi-scale integrals beyond one
  loop}}, \href{https://doi.org/10.1016/j.cpc.2015.05.022}{\emph{Comput. Phys.
  Commun.} {\bfseries 196} (2015) 470}
  [\href{https://arxiv.org/abs/1502.06595}{{\ttfamily 1502.06595}}].

\bibitem{Ronca:2019kcw}
J.~Ronca, \emph{{NNLO QED contribution to the $\mu e\to \mu e$ elastic
  scattering}},  in \emph{Flavour changing and conserving processes
  (FCCP2019)}, 2019, \href{https://arxiv.org/abs/1912.05397}{{\ttfamily
  1912.05397}}.

\bibitem{Kinoshita:KLN}
T.~{Kinoshita}, \emph{{Mass Singularities of Feynman Amplitudes}},
  \href{https://doi.org/10.1063/1.1724268}{\emph{Journal of Mathematical
  Physics} {\bfseries 3} (1962) 650}.

\bibitem{Kuraev:1985hb}
E.~A. Kuraev and V.~S. Fadin, \emph{{On Radiative Corrections to $e^+ e^-$
  Single Photon Annihilation at High-Energy}}, {\emph{Sov. J. Nucl. Phys.}
  {\bfseries 41} (1985) 466}.

\bibitem{Ellis:166310}
J.~R. Ellis and R.~D. Peccei, \emph{{Physics at LEP v1}},  in \emph{LEP Physics
  Jamboree}, (Geneva), CERN, CERN, 1986,
  \href{https://doi.org/10.5170/CERN-1986-002-V-1}{DOI}.

\bibitem{Frixione:2019lga}
S.~Frixione, \emph{{Initial conditions for electron and photon structure and
  fragmentation functions}},
  \href{https://doi.org/10.1007/JHEP11(2019)158}{\emph{JHEP} {\bfseries 11}
  (2019) 158} [\href{https://arxiv.org/abs/1909.03886}{{\ttfamily
  1909.03886}}].

\bibitem{Penin:2005kf}
A.~A. Penin, \emph{{Two-loop corrections to Bhabha scattering}},
  \href{https://doi.org/10.1103/PhysRevLett.95.010408}{\emph{Phys. Rev. Lett.}
  {\bfseries 95} (2005) 010408}
  [\href{https://arxiv.org/abs/hep-ph/0501120}{{\ttfamily hep-ph/0501120}}].

\bibitem{Penin:2005eh}
A.~A. Penin, \emph{{Two-loop photonic corrections to massive Bhabha
  scattering}},
  \href{https://doi.org/10.1016/j.nuclphysb.2005.11.016}{\emph{Nucl. Phys.}
  {\bfseries B734} (2006) 185}
  [\href{https://arxiv.org/abs/hep-ph/0508127}{{\ttfamily hep-ph/0508127}}].

\bibitem{Mitov:2006xs}
A.~Mitov and S.~Moch, \emph{{The Singular behavior of massive QCD amplitudes}},
  \href{https://doi.org/10.1088/1126-6708/2007/05/001}{\emph{JHEP} {\bfseries
  05} (2007) 001} [\href{https://arxiv.org/abs/hep-ph/0612149}{{\ttfamily
  hep-ph/0612149}}].

\bibitem{Becher:2007cu}
T.~Becher and K.~Melnikov, \emph{{Two-loop QED corrections to Bhabha
  scattering}},
  \href{https://doi.org/10.1088/1126-6708/2007/06/084}{\emph{JHEP} {\bfseries
  06} (2007) 084} [\href{https://arxiv.org/abs/0704.3582}{{\ttfamily
  0704.3582}}].

\bibitem{Engel:2018fsb}
T.~Engel, C.~Gnendiger, A.~Signer and Y.~Ulrich, \emph{{Small-mass effects in
  heavy-to-light form factors}},
  \href{https://doi.org/10.1007/JHEP02(2019)118}{\emph{JHEP} {\bfseries 02}
  (2019) 118} [\href{https://arxiv.org/abs/1811.06461}{{\ttfamily
  1811.06461}}].

\bibitem{Becher:2014oda}
T.~Becher, A.~Broggio and A.~Ferroglia, \emph{{Introduction to Soft-Collinear
  Effective Theory}},
  \href{https://doi.org/10.1007/978-3-319-14848-9}{\emph{Lect. Notes Phys.}
  {\bfseries 896} (2015) pp.1}
  [\href{https://arxiv.org/abs/1410.1892}{{\ttfamily 1410.1892}}].

\bibitem{Beneke:1997zp}
M.~Beneke and V.~A. Smirnov, \emph{{Asymptotic expansion of Feynman integrals
  near threshold}},
  \href{https://doi.org/10.1016/S0550-3213(98)00138-2}{\emph{Nucl.Phys.}
  {\bfseries B522} (1998) 321}
  [\href{https://arxiv.org/abs/hep-ph/9711391}{{\ttfamily hep-ph/9711391}}].

\bibitem{Fael:2019nsf}
M.~Fael and M.~Passera, \emph{{Muon-electron scattering at NNLO: the hadronic
  corrections}},
  \href{https://doi.org/10.1103/PhysRevLett.122.192001}{\emph{Phys. Rev. Lett.}
  {\bfseries 122} (2019) 192001}
  [\href{https://arxiv.org/abs/1901.03106}{{\ttfamily 1901.03106}}].

\bibitem{Yennie:1961ad}
D.~R. Yennie, S.~C. Frautschi and H.~Suura, \emph{{The infrared divergence
  phenomena and high-energy processes}},
  \href{https://doi.org/10.1016/0003-4916(61)90151-8}{\emph{Annals Phys.}
  {\bfseries 13} (1961) 379}.

\bibitem{Becher:2009kw}
T.~Becher and M.~Neubert, \emph{{Infrared singularities of QCD amplitudes with
  massive partons}}, \href{https://doi.org/10.1103/PhysRevD.79.125004,
  10.1103/PhysRevD.80.109901}{\emph{Phys. Rev.} {\bfseries D79} (2009) 125004}
  [\href{https://arxiv.org/abs/0904.1021}{{\ttfamily 0904.1021}}].

\bibitem{Ablinger:2018yae}
J.~Ablinger, J.~Bl{\"u}mlein, P.~Marquard, N.~Rana and C.~Schneider,
  \emph{{Heavy Quark Form Factors at Three Loops in the Planar Limit}},
  \href{https://doi.org/10.1016/j.physletb.2018.05.077}{\emph{Phys. Lett.}
  {\bfseries B782} (2018) 528}
  [\href{https://arxiv.org/abs/1804.07313}{{\ttfamily 1804.07313}}].

\bibitem{Blumlein:2019oas}
J.~Bl{\"u}mlein, P.~Marquard, N.~Rana and C.~Schneider, \emph{{The Heavy
  Fermion Contributions to the Massive Three Loop Form Factors}},
  \href{https://doi.org/10.1016/j.nuclphysb.2019.114751}{\emph{Nucl. Phys.}
  {\bfseries B949} (2019) 114751}
  [\href{https://arxiv.org/abs/1908.00357}{{\ttfamily 1908.00357}}].

\bibitem{Naterop:2019xaf}
L.~Naterop, A.~Signer and Y.~Ulrich, \emph{{$handyG$ - rapid numerical
  evaluation of generalised polylogarithms in Fortran}},
  \href{https://arxiv.org/abs/1909.01656}{{\ttfamily 1909.01656}}.

\bibitem{Broadhurst:1993mw}
D.~J. Broadhurst, J.~Fleischer and O.~V. Tarasov, \emph{{Two loop two point
  functions with masses: Asymptotic expansions and Taylor series, in any
  dimension}}, \href{https://doi.org/10.1007/BF01474625}{\emph{Z. Phys.}
  {\bfseries C60} (1993) 287}
  [\href{https://arxiv.org/abs/hep-ph/9304303}{{\ttfamily hep-ph/9304303}}].

\bibitem{Kuhn:1998ze}
J.~H. Kuhn and M.~Steinhauser, \emph{{A Theory driven analysis of the effective
  QED coupling at $M_Z$}},
  \href{https://doi.org/10.1016/S0370-2693(98)00908-3}{\emph{Phys. Lett.}
  {\bfseries B437} (1998) 425}
  [\href{https://arxiv.org/abs/hep-ph/9802241}{{\ttfamily hep-ph/9802241}}].

\bibitem{Steinhauser:1998rq}
M.~Steinhauser, \emph{{Leptonic contribution to the effective electromagnetic
  coupling constant up to three loops}},
  \href{https://doi.org/10.1016/S0370-2693(98)00503-6}{\emph{Phys. Lett.}
  {\bfseries B429} (1998) 158}
  [\href{https://arxiv.org/abs/hep-ph/9803313}{{\ttfamily hep-ph/9803313}}].

\bibitem{Sturm:2013uka}
C.~Sturm, \emph{{Leptonic contributions to the effective electromagnetic
  coupling at four-loop order in QED}},
  \href{https://doi.org/10.1016/j.nuclphysb.2013.06.009}{\emph{Nucl. Phys.}
  {\bfseries B874} (2013) 698}
  [\href{https://arxiv.org/abs/1305.0581}{{\ttfamily 1305.0581}}].

\bibitem{Jegerlehner:2017gek}
F.~Jegerlehner, \emph{{The Anomalous Magnetic Moment of the Muon}},
  \href{https://doi.org/10.1007/978-3-319-63577-4}{\emph{Springer Tracts Mod.
  Phys.} {\bfseries 274} (2017) pp.1}.

\bibitem{Hagiwara:2003da}
K.~Hagiwara, A.~D. Martin, D.~Nomura and T.~Teubner, \emph{{Predictions for
  $g-2$ of the muon and $\alpha_\mathrm{QED}(M^2_Z)$}},
  \href{https://doi.org/10.1103/PhysRevD.69.093003}{\emph{Phys. Rev.}
  {\bfseries D69} (2004) 093003}
  [\href{https://arxiv.org/abs/hep-ph/0312250}{{\ttfamily hep-ph/0312250}}].

\bibitem{Hagiwara:2006jt}
K.~Hagiwara, A.~D. Martin, D.~Nomura and T.~Teubner, \emph{{Improved
  predictions for $g-2$ of the muon and $\alpha_\mathrm{QED}(M^2_Z)$}},
  \href{https://doi.org/10.1016/j.physletb.2007.04.012}{\emph{Phys. Lett.}
  {\bfseries B649} (2007) 173}
  [\href{https://arxiv.org/abs/hep-ph/0611102}{{\ttfamily hep-ph/0611102}}].

\bibitem{Hagiwara:2011af}
K.~Hagiwara, R.~Liao, A.~D. Martin, D.~Nomura and T.~Teubner,
  \emph{{$(g-2)_\mu$ and $\alpha(M_Z^2)$ re-evaluated using new precise data}},
  \href{https://doi.org/10.1088/0954-3899/38/8/085003}{\emph{J. Phys.}
  {\bfseries G38} (2011) 085003}
  [\href{https://arxiv.org/abs/1105.3149}{{\ttfamily 1105.3149}}].

\bibitem{Keshavarzi:2018mgv}
A.~Keshavarzi, D.~Nomura and T.~Teubner, \emph{{Muon $g-2$ and $\alpha(M_Z^2)$:
  a new data-based analysis}},
  \href{https://doi.org/10.1103/PhysRevD.97.114025}{\emph{Phys. Rev.}
  {\bfseries D97} (2018) 114025}
  [\href{https://arxiv.org/abs/1802.02995}{{\ttfamily 1802.02995}}].

\bibitem{Actis:2010gg}
{\scshape Working Group on Radiative Corrections and Monte Carlo Generators for
  Low Energies} collaboration, S.~Actis et~al., \emph{{Quest for precision in
  hadronic cross sections at low energy: Monte Carlo tools vs. experimental
  data}}, \href{https://doi.org/10.1140/epjc/s10052-010-1251-4}{\emph{Eur.
  Phys. J.} {\bfseries C66} (2010) 585}
  [\href{https://arxiv.org/abs/0912.0749}{{\ttfamily 0912.0749}}].

\bibitem{Ignatov:2016}
F.~Ignatov, \emph{{{\tt VPLITE}}},  2016,
  \href{{https://cmd.inp.nsk.su/~ignatov/vpl}}{{https://cmd.inp.nsk.su/~ignatov/vpl}}.

\bibitem{Melnikov:2001uw}
K.~Melnikov, \emph{{On the theoretical uncertainties in the muon anomalous
  magnetic moment}},
  \href{https://doi.org/10.1142/S0217751X01005602}{\emph{Int. J. Mod. Phys.}
  {\bfseries A16} (2001) 4591}
  [\href{https://arxiv.org/abs/hep-ph/0105267}{{\ttfamily hep-ph/0105267}}].

\bibitem{Passera:2004bj}
M.~Passera, \emph{{The Standard model prediction of the muon anomalous magnetic
  moment}}, \href{https://doi.org/10.1088/0954-3899/31/5/R01}{\emph{J. Phys.}
  {\bfseries G31} (2005) R75}
  [\href{https://arxiv.org/abs/hep-ph/0411168}{{\ttfamily hep-ph/0411168}}].

\bibitem{vanRitbergen:1998hn}
T.~van Ritbergen and R.~G. Stuart, \emph{{Hadronic contributions to the muon
  lifetime}}, \href{https://doi.org/10.1016/S0370-2693(98)00895-8}{\emph{Phys.
  Lett.} {\bfseries B437} (1998) 201}
  [\href{https://arxiv.org/abs/hep-ph/9802341}{{\ttfamily hep-ph/9802341}}].

\bibitem{Davydychev:2000ee}
A.~I. Davydychev, K.~Schilcher and H.~Spiesberger, \emph{{Hadronic corrections
  at $\mathcal{O}(\alpha^2)$ to the energy spectrum of muon decay}},
  \href{https://doi.org/10.1007/s100520100577}{\emph{Eur. Phys. J.} {\bfseries
  C19} (2001) 99} [\href{https://arxiv.org/abs/hep-ph/0011221}{{\ttfamily
  hep-ph/0011221}}].

\bibitem{Actis:2007fs}
S.~Actis, M.~Czakon, J.~Gluza and T.~Riemann, \emph{{Virtual hadronic and
  leptonic contributions to Bhabha scattering}},
  \href{https://doi.org/10.1103/PhysRevLett.100.131602}{\emph{Phys. Rev. Lett.}
  {\bfseries 100} (2008) 131602}
  [\href{https://arxiv.org/abs/0711.3847}{{\ttfamily 0711.3847}}].

\bibitem{Kuhn:2008zs}
J.~H. Kuhn and S.~Uccirati, \emph{{Two-loop QED hadronic corrections to Bhabha
  scattering}},
  \href{https://doi.org/10.1016/j.nuclphysb.2008.08.002}{\emph{Nucl. Phys.}
  {\bfseries B806} (2009) 300}
  [\href{https://arxiv.org/abs/0807.1284}{{\ttfamily 0807.1284}}].

\bibitem{CarloniCalame:2011zq}
C.~Carloni~Calame, H.~Czyz, J.~Gluza, M.~Gunia, G.~Montagna, O.~Nicrosini
  et~al., \emph{{NNLO leptonic and hadronic corrections to Bhabha scattering
  and luminosity monitoring at meson factories}},
  \href{https://doi.org/10.1007/JHEP07(2011)126}{\emph{JHEP} {\bfseries 07}
  (2011) 126} [\href{https://arxiv.org/abs/1106.3178}{{\ttfamily 1106.3178}}].

\bibitem{Denner:2016kdg}
A.~Denner, S.~Dittmaier and L.~Hofer, \emph{{Collier: a fortran-based Complex
  One-Loop LIbrary in Extended Regularizations}},
  \href{https://doi.org/10.1016/j.cpc.2016.10.013}{\emph{Comput. Phys. Commun.}
  {\bfseries 212} (2017) 220}
  [\href{https://arxiv.org/abs/1604.06792}{{\ttfamily 1604.06792}}].

\bibitem{Patel:2015tea}
H.~H. Patel, \emph{{Package-X: A Mathematica package for the analytic
  calculation of one-loop integrals}},
  \href{https://doi.org/10.1016/j.cpc.2015.08.017}{\emph{Comput. Phys. Commun.}
  {\bfseries 197} (2015) 276}
  [\href{https://arxiv.org/abs/1503.01469}{{\ttfamily 1503.01469}}].

\bibitem{Fael:2018dmz}
M.~Fael, \emph{{Hadronic corrections to $\mu$-$e$ scattering at NNLO with
  space-like data}}, \href{https://doi.org/10.1007/JHEP02(2019)027}{\emph{JHEP}
  {\bfseries 02} (2019) 027}
  [\href{https://arxiv.org/abs/1808.08233}{{\ttfamily 1808.08233}}].

\bibitem{pagani2017}
L.~Pagani, \emph{{A new approach to muon $g-2$ with space-like data: analysis
  and fitting procedure.}},  {Tesi di Laurea Magistrale}, University of
  Bologna, 2017.

\bibitem{Beneke:2005}
M.~Beneke, \emph{Soft-collinear effective theory},  in \emph{Helmholtz
  International Summer School: Heavy Quark Physics}, Dubna, 2005.

\bibitem{Chiu:2011qc}
J.-y. Chiu, A.~Jain, D.~Neill and I.~Z. Rothstein, \emph{{The Rapidity
  Renormalization Group}},
  \href{https://doi.org/10.1103/PhysRevLett.108.151601}{\emph{Phys. Rev. Lett.}
  {\bfseries 108} (2012) 151601}
  [\href{https://arxiv.org/abs/1104.0881}{{\ttfamily 1104.0881}}].

\bibitem{Becher:2011pf}
T.~Becher, G.~Bell and M.~Neubert, \emph{{Factorization and Resummation for Jet
  Broadening}},
  \href{https://doi.org/10.1016/j.physletb.2011.09.005}{\emph{Phys. Lett.}
  {\bfseries B704} (2011) 276}
  [\href{https://arxiv.org/abs/1104.4108}{{\ttfamily 1104.4108}}].

\bibitem{Barberio:1993qi}
E.~Barberio and Z.~Was, \emph{{PHOTOS: A Universal Monte Carlo for QED
  radiative corrections. Version 2.0}},
  \href{https://doi.org/10.1016/0010-4655(94)90074-4}{\emph{Comput. Phys.
  Commun.} {\bfseries 79} (1994) 291}.

\bibitem{Jadach:1995nk}
S.~Jadach, W.~Placzek and B.~F.~L. Ward, \emph{{BHWIDE 1.00: $O(\alpha)$ YFS
  exponentiated Monte Carlo for Bhabha scattering at wide angles for LEP-1 /
  SLC and LEP-2}},
  \href{https://doi.org/10.1016/S0370-2693(96)01382-2}{\emph{Phys. Lett.}
  {\bfseries B390} (1997) 298}
  [\href{https://arxiv.org/abs/hep-ph/9608412}{{\ttfamily hep-ph/9608412}}].

\bibitem{Hamilton:2006xz}
K.~Hamilton and P.~Richardson, \emph{{Simulation of QED radiation in particle
  decays using the YFS formalism}},
  \href{https://doi.org/10.1088/1126-6708/2006/07/010}{\emph{JHEP} {\bfseries
  07} (2006) 010} [\href{https://arxiv.org/abs/hep-ph/0603034}{{\ttfamily
  hep-ph/0603034}}].

\bibitem{Schonherr:2008av}
M.~Schonherr and F.~Krauss, \emph{{Soft Photon Radiation in Particle Decays in
  SHERPA}}, \href{https://doi.org/10.1088/1126-6708/2008/12/018}{\emph{JHEP}
  {\bfseries 12} (2008) 018} [\href{https://arxiv.org/abs/0810.5071}{{\ttfamily
  0810.5071}}].

\bibitem{Krauss:2018djz}
F.~Krauss, J.~M. Lindert, R.~Linten and M.~Sch{\"o}nherr, \emph{{Accurate
  simulation of W, Z and Higgs boson decays in Sherpa}},
  \href{https://doi.org/10.1140/epjc/s10052-019-6614-x}{\emph{Eur. Phys. J.}
  {\bfseries C79} (2019) 143}
  [\href{https://arxiv.org/abs/1809.10650}{{\ttfamily 1809.10650}}].

\bibitem{Balossini:2006wc}
G.~Balossini, C.~M. Carloni~Calame, G.~Montagna, O.~Nicrosini and F.~Piccinini,
  \emph{{Matching perturbative and parton shower corrections to Bhabha process
  at flavour factories}},
  \href{https://doi.org/10.1016/j.nuclphysb.2006.09.022}{\emph{Nucl. Phys.}
  {\bfseries B758} (2006) 227}
  [\href{https://arxiv.org/abs/hep-ph/0607181}{{\ttfamily hep-ph/0607181}}].

\bibitem{Balossini:2008xr}
G.~Balossini, C.~Bignamini, C.~M. Carloni~Calame, G.~Montagna, O.~Nicrosini and
  F.~Piccinini, \emph{{Photon pair production at flavour factories with per
  mille accuracy}},
  \href{https://doi.org/10.1016/j.physletb.2008.04.007}{\emph{Phys. Lett.}
  {\bfseries B663} (2008) 209}
  [\href{https://arxiv.org/abs/0801.3360}{{\ttfamily 0801.3360}}].

\bibitem{CarloniCalame:2006zq}
C.~M. Carloni~Calame, G.~Montagna, O.~Nicrosini and A.~Vicini, \emph{{Precision
  electroweak calculation of the charged current Drell-Yan process}},
  \href{https://doi.org/10.1088/1126-6708/2006/12/016}{\emph{JHEP} {\bfseries
  12} (2006) 016} [\href{https://arxiv.org/abs/hep-ph/0609170}{{\ttfamily
  hep-ph/0609170}}].

\bibitem{CarloniCalame:2007cd}
C.~M. Carloni~Calame, G.~Montagna, O.~Nicrosini and A.~Vicini, \emph{{Precision
  electroweak calculation of the production of a high transverse-momentum
  lepton pair at hadron colliders}},
  \href{https://doi.org/10.1088/1126-6708/2007/10/109}{\emph{JHEP} {\bfseries
  10} (2007) 109} [\href{https://arxiv.org/abs/0710.1722}{{\ttfamily
  0710.1722}}].

\bibitem{Boselli:2015aha}
S.~Boselli, C.~M. Carloni~Calame, G.~Montagna, O.~Nicrosini and F.~Piccinini,
  \emph{{Higgs boson decay into four leptons at NLOPS electroweak accuracy}},
  \href{https://doi.org/10.1007/JHEP06(2015)023}{\emph{JHEP} {\bfseries 06}
  (2015) 023} [\href{https://arxiv.org/abs/1503.07394}{{\ttfamily
  1503.07394}}].

\bibitem{CarloniCalame:2001ny}
C.~M. Carloni~Calame, \emph{{An Improved parton shower algorithm in QED}},
  \href{https://doi.org/10.1016/S0370-2693(01)01108-X}{\emph{Phys.\ Lett.\ B}
  {\bfseries 520} (2001) 16}
  [\href{https://arxiv.org/abs/hep-ph/0103117}{{\ttfamily hep-ph/0103117}}].

\bibitem{Broggio:2014yca}
A.~Broggio, A.~S. Papanastasiou and A.~Signer, \emph{{Renormalization-group
  improved fully differential cross sections for top pair production}},
  \href{https://doi.org/10.1007/JHEP10(2014)098}{\emph{JHEP} {\bfseries 10}
  (2014) 098} [\href{https://arxiv.org/abs/1407.2532}{{\ttfamily 1407.2532}}].

\bibitem{Catani:1989et}
S.~Catani and L.~Trentadue, \emph{{Fermion pair exponentiation in QED}},
  {\emph{J. Exp. Theor. Phys. Lett.} {\bfseries 51} (1990) 83}.

\bibitem{Skrzypek:1992vk}
M.~Skrzypek, \emph{{Leading logarithmic calculations of QED corrections at
  LEP}}, {\emph{Acta Phys.\ Polon.\ B} {\bfseries 23} (1992) 135}.

\bibitem{Arbuzov:2010zzb}
A.~Arbuzov, V.~Bytev, E.~Kuraev, E.~Tomasi-Gustafsson and Y.~Bystritskiy,
  \emph{{Structure function approach in QED for high energy processes}},
  \href{https://doi.org/10.1134/S1063779610030020}{\emph{Phys.\ Part.\ Nucl.}
  {\bfseries 41} (2010) 394}.

\bibitem{Jegerlehner:2012}
F.~Jegerlehner, \emph{{{\tt hadr5n12}}},  2012,
  \url{http://www-com.phyisk.hu-berlin.de/~fjeger/software.html}

\bibitem{Arbuzov:1995qd}
A.~B. Arbuzov, V.~S. Fadin, E.~A. Kuraev, L.~N. Lipatov, N.~P. Merenkov and
  L.~Trentadue, \emph{{Small angle electron - positron scattering with a per
  mille accuracy}},
  \href{https://doi.org/10.1016/S0550-3213(96)00490-7}{\emph{Nucl. Phys.}
  {\bfseries B485} (1997) 457}
  [\href{https://arxiv.org/abs/hep-ph/9512344}{{\ttfamily hep-ph/9512344}}].

\bibitem{Arbuzov:1995vj}
A.~Arbuzov, E.~Kuraev, N.~Merenkov and L.~Trentadue, \emph{{Virtual and soft
  real pair production in large angle Bhabha scattering}}, {\emph{Phys. Atom.
  Nucl.} {\bfseries 60} (1997) 591}.

\bibitem{Arbuzov:1995cn}
A.~Arbuzov, E.~Kuraev, N.~Merenkov and L.~Trentadue, \emph{{Pair production in
  small angle Bhabha scattering}}, {\emph{J. Exp. Theor. Phys.} {\bfseries 81}
  (1995) 638} [\href{https://arxiv.org/abs/hep-ph/9509405}{{\ttfamily
  hep-ph/9509405}}].

\bibitem{Arbuzov:1995vi}
A.~Arbuzov, E.~Kuraev, N.~Merenkov and L.~Trentadue, \emph{{Hard pair
  production in large angle Bhabha scattering}},
  \href{https://doi.org/10.1016/0550-3213(96)00287-8}{\emph{Nucl. Phys.}
  {\bfseries B474} (1996) 271}.

\bibitem{Montagna:1998vb}
G.~Montagna, M.~Moretti, O.~Nicrosini, A.~Pallavicini and F.~Piccinini,
  \emph{{Light pair correction to Bhabha scattering at small angle}},
  \href{https://doi.org/10.1016/S0550-3213(99)00064-4}{\emph{Nucl. Phys.}
  {\bfseries B547} (1999) 39}
  [\href{https://arxiv.org/abs/hep-ph/9811436}{{\ttfamily hep-ph/9811436}}].

\end{thebibliography}\endgroup
}

\end{document}